\documentclass[journal,10pt]{IEEEtran}
\usepackage[cmex10]{amsmath}
\usepackage{graphicx}
\usepackage{algorithm}
\usepackage{algorithmic}
\usepackage{stfloats}
\usepackage{epstopdf}
\usepackage[numbers,sort&compress]{natbib}
\usepackage{multirow}
\usepackage{color}
\usepackage{balance}
\usepackage{amsmath, amssymb}
\usepackage{amsfonts}
\usepackage{caption}
\usepackage{subfigure}
\usepackage{hyperref}
\usepackage{comment}
\usepackage{bm}
\newcommand{\overbar}[1]{\mkern 1.5mu\overline{\mkern-1.5mu#1\mkern-1.5mu}\mkern 1.5mu}
\ifCLASSINFOpdf
\else
\fi
\begin{document}
\title{STAR-RIS-Assisted-Full-Duplex Jamming Design for Secure Wireless Communications System}
\author{Yun~Wen,~\IEEEmembership{Student Member,~IEEE,}
Gaojie~Chen,~\IEEEmembership{Senior Member,~IEEE,}
Sisai~Fang,~\IEEEmembership{Student Member,~IEEE,}
Zheng~Chu,~\IEEEmembership{Senior Member,~IEEE,}
Pei Xiao,~\IEEEmembership{Senior Member,~IEEE,}
and Rahim Tafazolli,~\IEEEmembership{Senior Member,~IEEE}

\thanks{\noindent Yun Wen, Gaojie Chen, Zheng Chu, Pei Xiao and Rahim Tafazolli are with the Institute for Communication Systems (ICS), 5GIC \& 6GIC, University of Surrey, Guildford, Surrey GU2 7XH, U.K. (e-mail: \{yun.wen, gaojie.chen, zheng.chu, p.xiao, r.tafazolli\}@surrey.ac.uk. (Corresponding author: Gaojie Chen.)}
\thanks{Sisai Fang is with the School of Engineering, University of Leicester, Leicester LE1 7RH, U.K. (e-mail: sf305@leicester.ac.uk).}
}

\maketitle

\begin{abstract}
Physical layer security (PLS) technologies are expected to play an important role in the next-generation wireless networks, by providing secure communication to protect critical and sensitive information from illegitimate devices. In this paper, we propose a novel secure communication scheme where the legitimate receiver use full-duplex (FD) technology to transmit jamming signals with the assistance of simultaneous transmitting and reflecting reconfigurable intelligent surface (STAR-RIS) which can operate under the energy splitting (ES) model and the mode switching (MS) model, to interfere with the undesired reception by the eavesdropper. We aim to maximize the secrecy capacity by jointly optimizing the FD beamforming vectors, amplitudes and phase shift coefficients for the ES-RIS, and mode selection and phase shift coefficients for the MS-RIS. With above optimization, the proposed scheme can concentrate the jamming signals on the eavesdropper while simultaneously eliminating the self-interference (SI) in the desired receiver. 
To tackle the coupling effect of multiple variables, we propose an alternating optimization algorithm to solve the problem iteratively. Furthermore,  we handle the non-convexity of the problem by the the successive convex approximation (SCA) scheme for the beamforming optimizations, amplitudes and phase shifts optimizations for the ES-RIS, as well as the phase shifts optimizations for the MS-RIS. In addition, we adopt a semi-definite relaxation (SDR) and Gaussian randomization process to overcome the difficulty introduced by the binary nature of mode optimization of the MS-RIS. Simulation results validate the performance of our proposed schemes as well as the efficacy of adapting both two types of STAR-RISs in enhancing secure communications when compared to the traditional self-interference cancellation technology. 
\end{abstract}


\IEEEpeerreviewmaketitle

\section{Introduction}
The rapid development of wireless communication technologies in recent years has accelerated the spread of wireless-connected applications into every aspect of our society. On the other hand, cybercrimes focused
on wireless communications, especially those applied to mission-critical services such as connected cars and telemedicine systems, may become realistic and cause new social problems \cite{Security1}. Due to its broadcast nature, wireless communication can be overheard by any undesired eavesdroppers in the communication range, thus is inherently more vulnerable in security compared to its wired counterpart. 

To provide robust security for wireless communications, cryptographic technologies have been widely adopted in current systems. However, the computational complexity based security is expected to face the challenge due to the development of quantum computing and is also difficult to be applied to devices with limited computation power such as those in the Internet of Things (IoT) networks \cite{Security2}. On the other hand, physical layer security (PLS) technologies exploit the physical characteristics of the wireless channel to achieve an information-theoretically secure communication, therefore have attracted much attention as a promising solution to safeguard the next generation wireless network \cite{PLS1}.  In recent years, numerical PLS technologies, such as secure beamforming, artificial noise aided transmission and channel state information (CSI)-based encryption, have been proposed to enhance the wireless security \cite{PLS_survey}.   

The full-duplex (FD) jamming scheme is considered as one of promising PLS technologies, where the legitimate receiver (Bob) or a relay receives information signals from the transmitter (Alice) and simultaneously transmits jamming signals to disturb the reception of the eavesdropper (Eve) \cite{gaojiepls}. The FD scheme can enhance secure communication in various scenarios, especially when the Eve is located near Bob and have a channel similar to the legitimate one between Alice and Bob. Its performance has been well investigated under numerous scenarios with perfect, imperfect, or even no channel state information (CSI) of Eve \cite{FD1, FD2, FD3}. Despite its outstanding performance, FD secure communication has to tackle the inherent self-interference (SI) problem. The SI imposed by Bob’s transmission causes a non-negligible effect on its own signal reception, thus has to be mitigated with proper canceling techniques to achieve a positive secrecy capacity \cite{FD_SI}. Numerous self-interference cancellation (SIC) techniques \cite{SIC_new1,BeamBased,SIC_new3,SIC_new4} have been developed which can even reduce the SI power to the noise floor. However, due to their high hardware cost by introducing extra circuit between antennas, their application are limited to those with small number of antennas, thus can not provide efficient secure degrees of freedom (DoF) to disturb the Eve. For FD communications with a large number of antennas, SIC techniques such as hybrid beamforming (HYBF) have also been proposed \cite{HYBF2} with reduced number of radio frequency (RF) chains. Although these technologies have the potential of improving secure DoF, they still suffer high hardware complexity and power consumption \cite{SIC_new2}. 

To provide extra DoF with low complexity and power consumption, the reconfigurable intelligent surface (RIS) has been proposed and attracted increasing attention from both academia and industry in recent years \cite{Cognitive}. Consisting of a large number of passive reflecting components with software-controlled amplitudes and phases, RIS can provide a programmable propagation environment to enhance the performance of wireless communication at a low cost. Due to its significant potential, RIS-assisted communication has been considered a promising enabling technology in various areas including the millimeter-wave (mmWave) communication and intelligent transportation systems \cite{magRIS}. The benefit of merging RIS with PLS technologies has also been studied, from different perspectives such as enhancing the desired signals and degrading the eavesdropping capacity \cite{RIS-sec1,RIS-sec2}. In addition, some latest works on PLS studied the synergistic effect of combining RIS and FD communication \cite{RIS-FD-sec1,RIS-FD-sec2}, in order to reap the advantages of both technologies to further enhance secure communication. However, these works impose the unrealistic assumption of perfect SI cancellation, and fail to address the problem of high hardware complexity and power consumption incroduced by FD communications.

Recently, simultaneous transmitting and reflecting (STAR) RIS, also known as intelligent
omni surface (IOS), has been developed as an update of traditional RISs \cite{NTT,STAR-1}. STAR-RIS's elements have dual functionalities of reflecting and refracting the received signals, which makes it possible to serve devices on both sides of the surface and consequently is more efficient in enhancing the radio coverage than traditional RISs \cite{IOS_cov}. The STAR-RIS was firstly introduced in \cite{Sisai_preprint} to solve the SI problem in FD systems with a low cost and power consumption, by adopting a STAR-RIS integrated transmitter whose SI can be eliminated with the reflection of STAR-RIS. However, to the best of our knowledge, how to adopt STAR-RIS assisted FD communications to enhance security and simultaneously cancel the SI has not been studied.

In this paper, we propose a novel STAR-RIS assisted FD secure communication scheme, to maximize the secrecy capacity by enhancing the FD jamming signal to disturb Eve's reception while simultaneously eliminating the SI from FD jamming. 
Optimization algorithms with two modes of the STAR-RIS, namely, energy splitting (ES)-RIS and mode switching (MS)-RIS, are studied and evaluated in this work. The main contributions of this paper are summarized as follows:

\begin{enumerate}
\item We are the first to adopt STAR-RIS into FD secure communications, which can not only reduce the SI to the same level as the traditional SIC technologies, but also control the jamming power received in Eve to achieve a significant improvement in secrecy capacity.  
\item We investigate a secrecy capacity maximization problem, where the beamforming vectors in both transmitter and receiver antennas at the legitimate users, the amplitudes and phase shifts for the ES-RIS or the phase shifts and mode selection for the MS-RIS are jointly optimized subject to transmit power and phase shifts constraints. 
\item We develop an alternating optimization-based algorithm to solve the coupling problem of multiple variables. The successive convex approximation (SCA) method is applied to tackle the non-convexity problem of the optimizations of beamforming, amplitudes and phase shifts for the ES-RIS, and phase shifts for the MS-RIS. In addition, a semi-definite relaxation (SDR) and Gaussian randomization scheme is adopted to address the binary optimizations of mode selection for the MS-RIS.
\item Simulation results validate the performance of the proposed scheme in improving the secrecy capacity, in comparison with traditional SIC scheme and without FD jamming. More specifically, we prove that with both the ES-RIS and MS-RIS the SI power can be reduced to the noise floor with a small number of RIS elements. On the other hand, the proposed algorithm efficiently enhances the jamming power in Eve and outperforms the traditional SIC scheme, thus achieves a cost-effective FD secure communication with the assist of low cost passive elements.
\end{enumerate}

The remainder of this paper is organized as follows. Section II describes the system model of the proposed STAR-RIS assisted FD secure communication system, followed by the problem formulation of the maximization of secrecy capacity. Section III illustrates the details of the proposed iterative algorithm of optimizing the beamforming vectors of the transmitter and the receiver, the amplitudes and phase shifts for the ES-RIS and the mode selection and phase shifts for the MS-RIS. Section IV presents numerical results to evaluate the efficiency of the proposed algorithm in comparison with benchmark schemes. We conclude this paper in Section VI. 

\textit{Notations:} In this paper, vectors and matrices are denoted by bold lowercase letters and bold uppercase letters, respectively. $\mathbf{a}^H$ gives the Hermitian of the vector $\mathbf{a}$, $\mathbf{a}^T$ and $\mathbf{a}^*$ denote its transpose and conjugate operators, respectively. $\left[\mathbf{A} \right]_{i, j}$ is the element located on the $i$th row and $j$th column of matrix $\mathbf{A}$ and ${\left[ {\mathbf{a}} \right]_i}$ is the $i$th element of vector ${\mathbf{a}}$. ${\mathbf{B}} \odot {{\mathbf{C}}}$ is the Hadamard product of ${\mathbf{B}}$ and ${\mathbf{C}}$. The trace of a matrix $\mathbf{A}$ is represented by $\text{Tr}(\mathbf{A})$. ${\mathbb{C}^{a \times b}}$ and ${\mathbb{R}^{a \times b}}$ are the complex matrix space of $a \times b$ and the real matrix space of $a \times b$, respectively.  $\operatorname{Re} \left\{  \cdot  \right\}$ and $|  \cdot  |$ denote the real part and the modulus of a complex value, respectively. ${\left\| \cdot \right\|}$ represents the Euclidean norm of a vector, $\mathbf{I}_n$ is a $n\times n$ identity matrix and $\textbf{1}_n$ stands for a $n\times 1$ identity vector. The notation $\left[x\right]^{+}$ denotes the operation of  $\max(x,0)$. ${\text{diag}}\left\{  \cdot  \right\}$ and ${\left(  \cdot  \right)^{\star}}$ denote the operator for diagonalization and the optimal value, respectively.  $\text{sgn}\left(  \cdot  \right)$ is the sign function of a real number and $\mathcal{O}\left(  \cdot  \right)$ is the big-O notation.

\section{System Model and Problem Formulation}

\subsection{System model}

\begin{figure*}
	\centering
	\subfigure[]{
		\begin{minipage}[b]{0.48\textwidth}
			\includegraphics[width=1\textwidth]{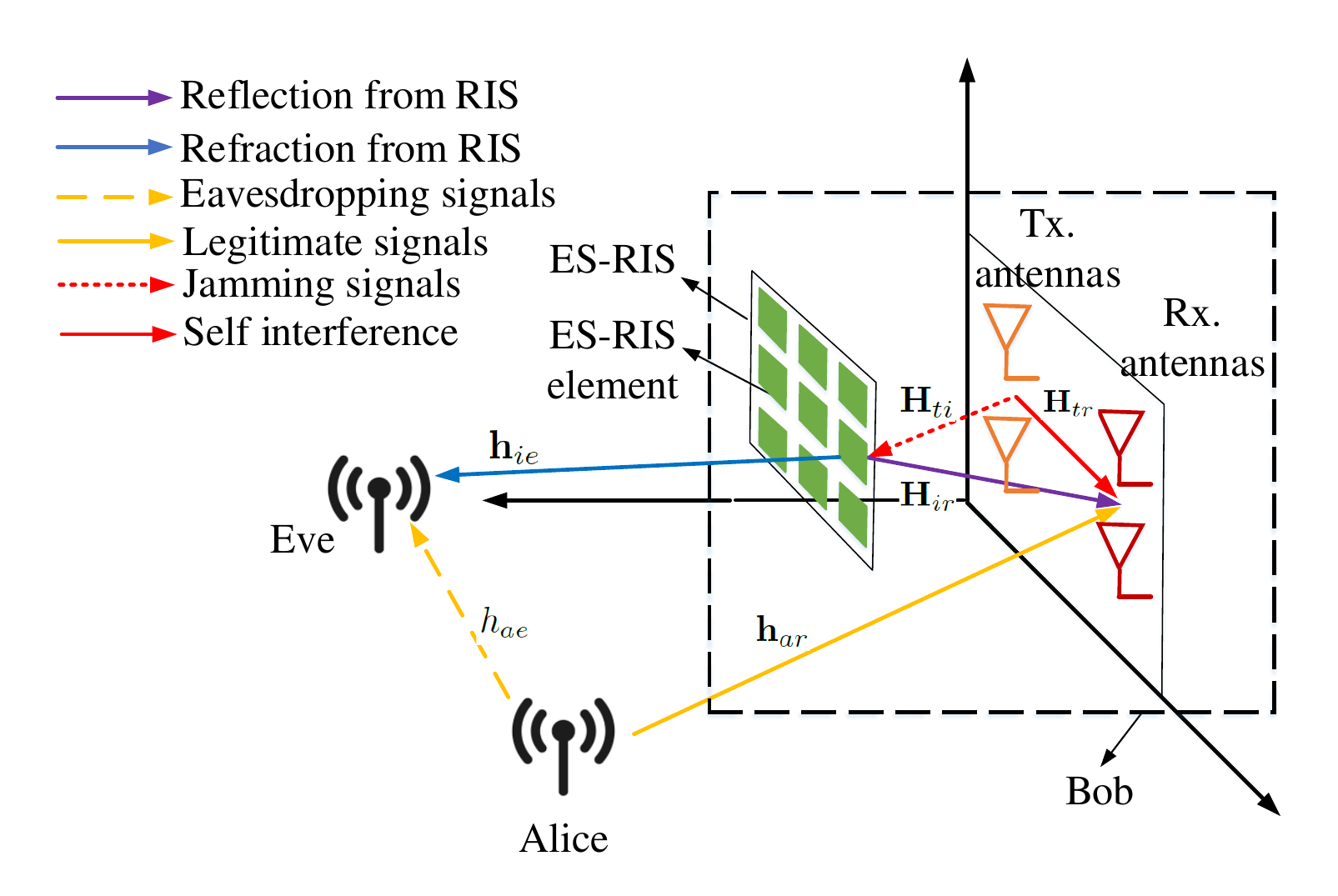}
		\end{minipage}
		\label{3D}
	}
    	\subfigure[]{
    		\begin{minipage}[b]{0.48\textwidth}
   		 	\includegraphics[width=1\textwidth]{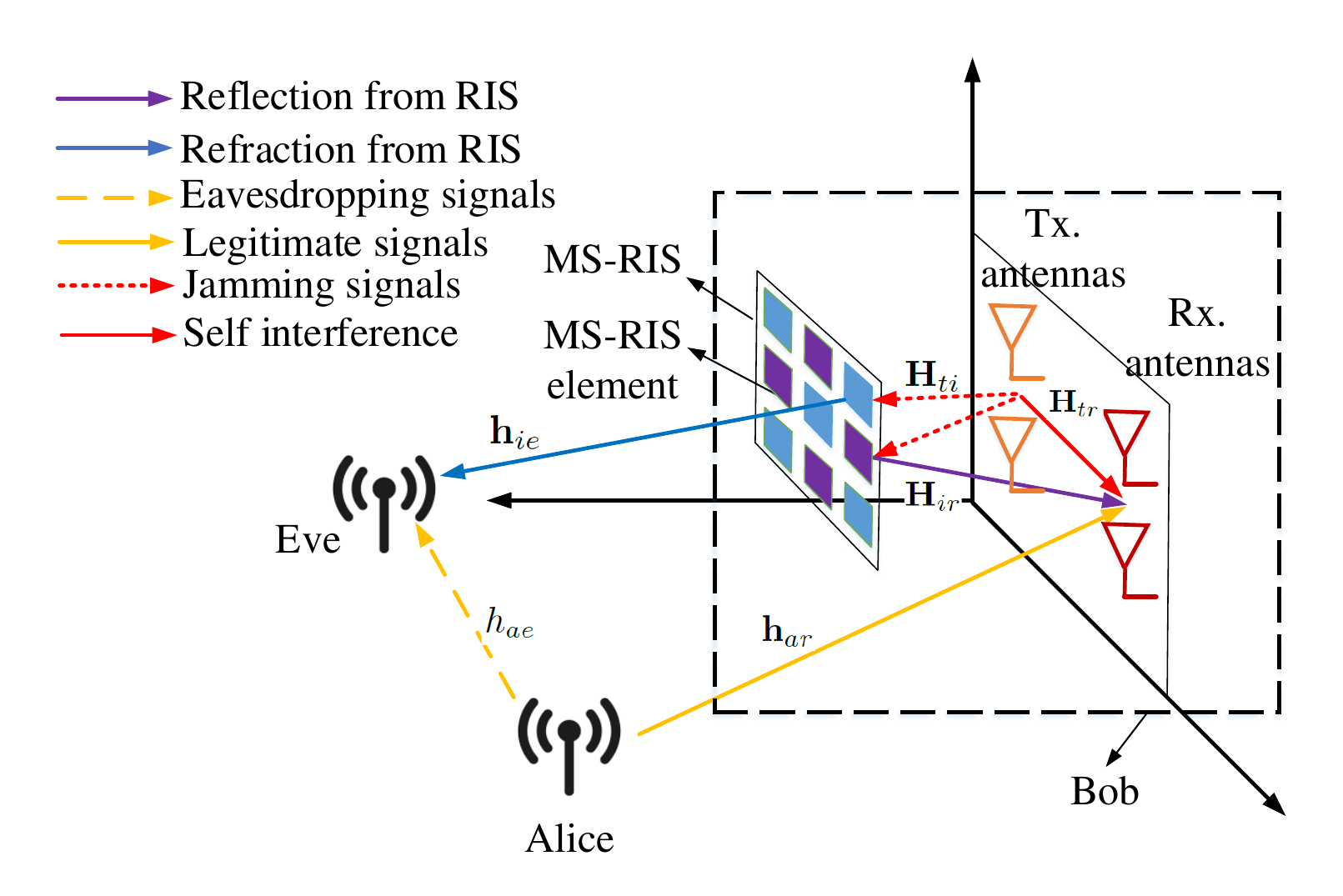}
    		\end{minipage}
		\label{3D_2}
    	}
	\caption{(a) An ES-RIS assisted FD network.  (b) A MS-RIS assisted FD network. }
	\label{fig:hor_2figs_1cap_2subcap}
\end{figure*}

We consider a STAR-RIS assisted secure communication system which consists of a transmitter Alice, a legitimate receiver Bob and an eavesdropper Eve. Since Eve can overhear the signals transmitted from Alice, Bob operates in the FD mode transmitting jamming signals when receiving the signals from Alice, to degrade the capacity of the Alice-Eve link to ensure secure communication. A $L$-element STAR-RIS is adopted to assist the jamming transmission of Bob, whose phase shift pattern is designed to maximize the power of jamming signals received in Eve and simultaneously mitigate the SI in the receive antennas of Bob. Here, we assume that Alice and Eve are equipped with a single antenna and Bob is equipped with $M$ transmit antennas integrated with a STAR-RIS and $N$ receive antennas. Furthermore, two types of STAR-RISs are considered in our model as in \cite{STAR-1}: 1) ES-RIS whose elements can reflect and refract signals simultaneously; 2) MS-RIS whose elements can only choose to reflect or refract signals at a certain time slot, as illustrated in Fig. \ref{3D} and Fig. \ref{3D_2}, respectively.

To define the effect of STAR-RIS on the transmitted jamming signals, we represent the reflecting and refracting coefficient matrices of the ES-RIS as ${\mathbf{R}_s}={\text{diag}}\left\{ {{u_{s,1}}{e^{j{\mu _{s,1}}}},{u_{s,2}}{e^{j{\mu _{s,2}}}}, \ldots ,{u_{s,L}}{e^{j{\mu _{s,L}}}}} \right\} \in {\mathbb{C}^{L \times L}}$ and ${{\mathbf{T}}_s}={\text{diag}}\left\{ {{v_{s,1}}{e^{j{\nu _{s,1}}}},{v_{s,2}}{e^{j{\nu _{s,2}}}}, \ldots ,{v_{s,L}}{e^{j{\nu _{s,L}}}}} \right\}\in{\mathbb{C}^{L \times L}}$, respectively. Here, $u_{s,l}$, $v_{s,l}$ denote the reflecting and refracting amplitudes at the $l$th element, and $\mu_{s,l}$, $\nu_{s,l}$ indicate the reflecting and refracting phase shifts of the $l$th element, respectively. In the same manner, corresponding matrices of the MS-RIS are represented as ${{\mathbf{R}}_m}={\text{diag}}\left\{ {{u_{m,1}}{e^{j{\mu _{m,1}}}},{u_{m,2}}{e^{j{\mu _{m,2}}}}, \ldots ,{u_{m,L}}{e^{j{\mu _{m,L}}}}} \right\} \in {\mathbb{C}^{L \times L}}$ and ${{\mathbf{T}}_m}={\text{diag}}\left\{ {{v_{m,1}}{e^{j{\nu _{m,1}}}},{v_{m,2}}{e^{j{\nu _{m,2}}}}, \ldots ,{v_{m,L}}{e^{j{\nu _{m,L}}}}} \right\} \in {\mathbb{C}^{L \times L}}$ 

According to \cite{STAR-2}, the constraints for the ES-RIS can be defined as
\begin{subequations}\label{constraint_1}
\begin{equation}\label{1a}
u_{s,l}^2 + v_{s,l}^2 \leq 1,
\end{equation}
\begin{equation}\label{1b}
0 \leq{u_{s,l}},{v_{s,l}} \leq 1,
\end{equation}
\begin{equation}\label{1c}
0 \leq {\mu _{s,l}},{\nu _{s,l}} < 2\pi , \ \  \forall l,
\end{equation}
\end{subequations}
\noindent where \eqref{1a} and \eqref{1b} indicate the limitation on the total power of reflection and refraction, and the amplitude limitation of either operation, respectively. 

On the other hand, the constraints for the MS-RIS are listed as follows:
\begin{subequations}\label{constraint_2}
\begin{equation}\label{2a}
{u_{m,l}} + {v_{m,l}} = 1,
\end{equation}
\begin{equation}\label{2b}
{u_{m,l}},{v_{m,l}} \in \left\{ {0,1} \right\},
\end{equation}
\begin{equation}\label{2c}
0 \leq {\mu _{m,l}},{\nu _{m,l}} < 2\pi,  \ \  \forall l,
\end{equation}
\end{subequations}
\noindent where \eqref{2a} and \eqref{2b} indicate that each element of the MS-RIS operates either in the reflection or refraction mode. 

The vector of jamming signals transmitted from Bob is represented as
\begin{equation}
{\mathbf{x_{t}}} = {\sqrt{P_{B}}\mathbf{w}s_{j}},
\end{equation}
where $P_{B}$ is the transmit power of Bob, $\mathbf{w}\in {\mathbb{C}^{M \times 1}}$ is the transmit beamforming vector where ${\text{Tr}}\left( {{\mathbf{w}}{{\mathbf{w}}^H}} \right) \leq 1$, and $s_{j}$ is the jamming signal. 

Then, the received signals in Bob for both the ES and MS modes can be expressed as 
\begin{equation}\label{signal_B}
{{\mathbf{y}}_{r,o}} =  {{\mathbf{h}}_{ar}^H}{\sqrt{P_{A}}s_{a}} \!+\! \left( {{\mathbf{H}}_{tr} + {\mathbf{H}}_{ir}^H{{\mathbf{R }}_o}{{\mathbf{H}}_{ti}}} \right){\sqrt{P_{B}}\mathbf{w}s_{j}} + {{\mathbf{n}}_r},
\end{equation}

\noindent where $o \in \left\{ {s,m} \right\}$ denotes the operation mode with either the ES-RIS or MS-RIS.  ${{\mathbf{h}}_{ar}} \in {\mathbb{C}^{1 \times N}}$ is the channel coefficients between Alice and the receive antennas of Bob. $P_{A}$ is the transmit power of Alice and $s_{a}$ is the information signal. ${{\mathbf{H}}_{tr}} \in {\mathbb{C}^{N \times M}}$, ${{\mathbf{H}}_{ir}} \in {\mathbb{C}^{L \times N}}$ and ${{\mathbf{H}}_{ti}} \in {\mathbb{C}^{L \times M}}$ are the matrices, the elements of which represent the coefficients of the channels between the transmit and the receive antennas of Bob, those between RIS elements and the receive antennas of Bob, and those between the transmit antennas of Bob and RIS elements, respectively. The second term in \eqref{signal_B} expresses the SI signals from the jamming transmission of Bob through both the direct path and reflection path from RIS. ${\mathbf{n}_r} \sim \mathcal{CN}\left( {0,\sigma _r^2{\mathbf{I}_N}} \right)$ represents the additive white Gaussian noise (AWGN) at the receive antennas of Bob and $\sigma _r^2$ is the noise power.

On the other hand, the received signals\footnote {We assume that the perfect CSI of all links including those with Eve are available for the optimization, because Eve may also be a user in the system whose CSI is possible to be estimated. The details of channel estimation is out of the scope of this paper but presented in \cite{CSI1, CSI2}. Since the proposed algorithm with perfect CSI gives a upper limit of secrecy performance, we also provide results with imperfect CSI in Section IV as a benchmark.}at Eve with the ES-RIS or MS-RIS is given by
\begin{equation}\label{signal_E}
{{y}_{e,o}} = {{h}}_{ae}^H{\sqrt{P_{A}}s_{a}} + {\mathbf{h}}_{ie}^H{{\mathbf{T}}_o}{{\mathbf{H}}_{ti}}{\sqrt{P_{B}}\mathbf{w}s_{j}} + {n}_e,
\end{equation}

\noindent where ${{{h}}_{ae}} \in {\mathbb{C}^{1 \times 1}}$ is the channel coefficient between Alice and Eve. The second term in \eqref{signal_E} is the received jamming signals from Bob through the refraction of RIS, where ${{\mathbf{h}}_{ie}} \in {\mathbb{C}^{L \times 1}}$ is the channel coefficients between the RIS elements and Eve's antenna. ${n}_e \sim \mathcal{CN}\left( {0,\sigma _e^2} \right)$ represents the AWGN in Eve and $\sigma _e^2$ is the noise power.

We assume the channel between the transmit antennas of Bob and RIS, the channel between the receive antennas of Bob and RIS and the channels between RIS elements and Eve as Rician channels where the line-of-sight (LOS) path is dominant in the propagation \cite{IOS_cov}. To derive the above mentioned channel coefficients, we characterize the relative position of the $m$th transmit antennas to the $l$th element of RIS by $\left( {{r_{l,m}},\theta _{l,m},\phi _{l,m}} \right)$. Here, we take the position of the $l$th element as the respective origin, thus ${r_{l,m}} \geq 0$, $\theta _{l,m} \in \left[ {0,\frac{\pi }{2}} \right]$ and $\phi _{l,m} \in \left[ {0,2\pi } \right]$ are the relative distance, the elevation angle and the azimuth angle of the $m$th transmit antenna, respectively. In the same manner, we define the relative positions of the receive antennas at the elements of the RIS, the positions of the receive antennas at the transmit antennas of Bob, and the positions of Alice's transmit antenna at the receive antennas of Bob as $\left( {{r_{l,n}},{\theta _{l,n}},{\phi _{l,n}}} \right)$, $\left( {{r_{m,n}},{\theta _{m,n}},{\phi _{m,n}}} \right)$ and $\left( {{r_{n,a}},{\theta _{n,a}},{\phi _{n,a}}} \right)$. With above definitions, we can derive the corresponding channel coefficients as follows:
\begin{equation}
\begin{gathered}
  {{\mathbf{H}}_{tr}} = \left[ {\frac{{\lambda \sqrt {{G^t}\left( {{\theta _{m,n}},{\phi _{m,n}}} \right){G^r}\left( {{\theta _{n,m}},{\phi _{n,m}}} \right)} }}{{4\pi r_{m,n}^{{\kappa  \mathord{\left/
 {\vphantom {\kappa  2}} \right.
 \kern-\nulldelimiterspace} 2}}}}} \right. \hfill \\
  {\left. { \qquad\quad \times \left( {\sqrt {\frac{K}{{K + 1}}} {e^{ - j\frac{{2\pi {r_{m,n}}}}{\lambda }}} \!+\! \sqrt {\frac{1}{{K + 1}}} h_{tr}^{nlos}} \right)} \right]_{n,m}}, \hfill \\ 
\end{gathered} 
\end{equation}

\begin{align}
&{{\mathbf{H}}_{ir}} = {\left[ {\frac{{\lambda \sqrt {{G^r}\left( {{\theta _{l,n}},{\phi _{l,n}}} \right)} }}{{4\pi {r_{l,n}}}}{e^{ - j\frac{{2\pi {r_{l,n}}}}{\lambda }}}} \right]_{l,n}}, \\
&{{\mathbf{H}}_{ti}} = {\left[ {\frac{{\lambda \sqrt {{G^t}\left( {{\theta _{l,m}},{\phi _{l,m}}} \right)} }}{{4\pi {r_{l,m}}}}{e^{ - j\frac{{2\pi {r_{l,m}}}}{\lambda }}}} \right]_{l,m}},
\end{align}

\begin{equation}
{{\mathbf{h}}_{ie}} \!=\! {\left[ {\frac{{\lambda }}{{4\pi r_{l,e}^{{\kappa  \mathord{\left/
 {\vphantom {\kappa  2}} \right.
 \kern-\nulldelimiterspace} 2}}}}\left( {\sqrt {\frac{K}{{K\!\! +\!\! 1}}} {e^{ -\! j\frac{{2\pi {r_{l,e}}}}{\lambda }}} \!+\! \sqrt {\frac{1}{{K \!+\!\! 1}}} h_{ie}^{nlos}} \right)} \right]_{l}},
\end{equation}

\noindent where $\lambda $ is the wavelength, ${G^t}\left( {\theta ,\phi } \right)$ and ${G^r}\left( {\theta ,\phi } \right)$ are the gains of Bob's transmit and receive antennas, respectively. $r_{m,n}$, $r_{l,n}$, $r_{l,m}$ and $r_{l,e}$ is the distance between the $m$th transmit antenna and the $n$th receive antenna of Bob, the distance between the $l$th element of the RIS and the $m$th transmit antenna of Bob, the distance between the $l$th element of the RIS and the $n$th receive antenna of Bob, and the distance between the $l$th element of the RIS and Eve, respectively. In addition, $K$ and $\kappa$ are the Rician factor and path loss exponent of the channels, while $h_{tr}^{nlos}$ and $h_{ie}^{nlos}$ are the non line-of-sight (NLOS) channel components, generalized using zero-mean and unit-variance circularly symmetric complex Gaussian random variables.

On the other hand, we assume the channel between Alice and Eve and the channels between Alice and the receive antennas of Bob are Rayleigh fading channels. Therefore, the corresponding channel coefficients can be expressed as ${{\mathbf{h}}_{ar}} = {\left[ {h_{f,n}/{\sqrt{r^{\alpha}_{n,a}}}} \right]_{n}}$ and ${h_{ae}} =  {h_{f,e}/\sqrt{r_{e,a}^\alpha}} $, respectively. Here $r_{n,e}$ and $r_{a,e}$ are the distance between the $n$th receive antenna of Bob and Eve, and the distance between Alice and Eve, respectively. $\alpha \in [2,6]$ is the path loss exponent depending on the frequency and surrounding environment. $h_{f,n},h_{f,e} \sim \mathcal{CN}\left( {0,\sigma_f^2} \right)$ are the channel components incorporating the small-scale Rayleigh fading variance $\sigma_f$.

We can thus represent the data rate of Bob and Eve as follows:
\begin{align}
&{R_{r,o}} = {\log _2}\left( {1 + \frac{{{{P_{A}\mathbf{r}\mathbf{h}}_{ar}^H}{\mathbf{h}}_{ar}}\mathbf{r}^H}{{P_{si}}+{\sigma _r^2}}} \right), \\
&{R_{e,o}} = {\log _2}\left( {1 + \frac{P_{A}{{{h}_{ae}}{h}_{ae}^H}}{{P_{j}}+{\sigma _e^2}}} \right),
\end{align}

\noindent where ${\mathbf{r}} \in {\mathbb{C}^{1 \times N}}$ is the receive beamforming vector in Bob. $P_{si}$ and $P_{j}$ are the SI power in the receive antennas of Bob and the power of jamming signals received in Eve, which can be given as 
\begin{align}
&{P_{si}} = {P_{B}\left\| {\mathbf{r}{{\mathbf{H}}_{r,o}}{\mathbf{w}}{{\mathbf{w}}^H}{\mathbf{H}}_{r,o}^H\mathbf{r}^H} \right\|},\\
&{P_{j}} = {P_{B}\left\| {{{\mathbf{h}}_{e,o}}{\mathbf{w}}{{\mathbf{w}}^H}{\mathbf{h}}_{e,o}^H} \right\|},
\end{align}

\noindent where ${{\mathbf{H}}_{r,o}} = {\mathbf{H}}_{tr} + {\mathbf{H}}_{ir}^H{\mathbf{R}_o}{{\mathbf{H}}_{ti}}$ and ${{\mathbf{h}}_{e,o}} = {\mathbf{h}}_{ie}^H{\mathbf{T }_o}{{\mathbf{H}}_{ti}}$. 

\subsection{Problem formulation}
In this paper, we aim to solve the optimization problem of maximizing secrecy capacity with STAR-RIS operating in either the ES or MS mode, subject to the power constraints and the phase shift unit modulus constraints. 
\begin{small}\begin{align}
\mathop {\max }\limits_{\mathbf{w},\mathbf{r},\mathbf{R}_o,\mathbf{T}_o}&\ \ {C_{o}}  \label{P1_OF}\\
{\text{s.t.}}\ \ & \ \   {\text{Tr}}\left( {{\mathbf{w}}{{\mathbf{w}}^H}} \right) \leq 1, \tag{\ref{P1_OF}{a}}  \label{limit_w} \\
& \ \   {\text{Tr}}\left( {{\mathbf{r}}{{\mathbf{r}}^H}} \right) \leq 1,  \label{limit_r} \tag{\ref{P1_OF}{b}}   \\
& \ \  P_{B} \left\| {{{\mathbf{H}}_{r,o}}{\mathbf{w}}{{\mathbf{w}}^H}{\mathbf{H}}_{r,o}^H} \right\| \leq {P_{th}}, \tag{\ref{P1_OF}{c}}  \label{SI_limit}\\
& \ \  \eqref{1a}-\eqref{1c} \ \text{or} \: \eqref{2a}- \eqref{2c}, \tag{\ref{P1_OF}{d}}  \label{P1_modulus}
\end{align}\end{small}where $C_{o}=\left[R_{r,o} - R_{e,o}\right]^{+}$ is the secrecy capacity\footnote{Although the secrecy capacity is generally expressed as $C_{o} = \left[R_{r,o} - R_{e,o}\right]^{+}$, we here adopt $C_{o} = R_{r,o} - R_{e,o}$ to simplify the optimization since the maximization operation with $0$ can be applied subsequently.}. \eqref{limit_w} and \eqref{limit_w} denote the power constraint of the beamforming vectors in the transmit and the receive antennas of Bob, respectively. The constraint of \eqref{SI_limit} indicates that the received power of SI in received antennas should be lower than a certain threshold $P_{th}$ to enable the subsequent signal processing in the analog-to-digital converter (ADC) \cite{ADC2}. In addition, the constraint in \eqref{P1_modulus} is chosen based on the mode of STAR-RIS. 

\section{Optimizations of Secrecy Capacity}
We notice that the problem defined in \eqref{P1_OF} involves joint optimization of multiple variables which are all coupled with each other, thus is difficult to be solved directly. Therefore, we here propose an algorithm based on the alternating optimization framework, where each variable is optimized separately by setting other variables to fixed values. By performing the alternating optimization for each variable sequentially and iteratively, the secrecy capacity will increase and converge to the optimal value eventually. To effectively utilize the iterative nature of alternating optimization, the SCA scheme is adopted to address the non-convexity of the optimization problem for each variable, where the first order Taylor expansion at the value obtained in previous iteration is adopted to approximate the objective function and constraints. In addition, we introduce SDR and the Gaussian randomization process to tackle the binary mode-selection problem for the MS mode which is NP-hard to be solved directly.

In the sequel, we will explain details of alternating optimization for each variable as well as the overall agorithm.

\subsection{Optimizing transmit beamforming vector $\mathbf{w}$}
Firstly, with given $\mathbf{r}$, $\mathbf{R}_o$ and $\mathbf{T}_o$, the optimization of beamforming vector $\mathbf{w}$ can be recast from \eqref{P1_OF} as 
\begin{small}\begin{align}
\mathop {\max }\limits_{\mathbf{w}}&\ \ {\log _2 \left(1 + \frac{A}{{P_{B}\left\| {\mathbf{r}{{\mathbf{H}}_{r,o}}{\mathbf{w}}{{\mathbf{w}}^H}{\mathbf{H}}_{r,o}^H\mathbf{r}^H} \right\|} + {\sigma _r^2}} \right)} \nonumber \\
&- {\log _2 \left(1 + \frac{B}{{P_{B}\left\| {{{\mathbf{h}}_{e,o}}{\mathbf{w}}{{\mathbf{w}}^H}{\mathbf{h}}_{e,o}^H} \right\|} + {\sigma _e^2}} \right)}  \label{P2_OF}\\
{\text{s.t.}}\ \ & \ \  {\text{Tr}}\left( {{\mathbf{w}}{{\mathbf{w}}^H}} \right) \leq 1,  \tag{\ref{P2_OF}{a}}  \label{P2_OF1} \\
& \ \ P_{B}{{\mathbf{w}}^H}{\left( {\mathbf{H}_{r,o}^H}{\mathbf{H}}_{r,o}\right)}{\mathbf{w}} \leq P_{th}, \tag{\ref{P2_OF}{b}} \label{P2_OF3}
\end{align}\end{small}

\noindent where $A = {{{P_{A}\mathbf{r}\mathbf{h}}_{ar}^H}{\mathbf{h}}_{ar}}\mathbf{r}^H$ and $B = P_{A}{{{\mathbf{h}}_{ae}}{\mathbf{h}}_{ae}^H}$. 

Since the objective function in \eqref{P2_OF} is non-concave with respect to $\mathbf{w}$, we transform it by introducing slack variables $\tau_w$, $\xi_w$ and $\varrho_w$ as follows:
\begin{small}
\begin{align}
\mathop {\max }\limits_{\mathbf{w},\xi_w, \tau_w, \varrho_w}&\ \ {\log _2 \left(1 + \frac{A}{\tau_w} \right)} 
- \xi_w  \label{P3_OF} \\
{\text{s.t.}}\ \ & \ \  \tau_w \geq {P_{B}\left\| {\mathbf{r}{{\mathbf{H}}_{r,o}}{\mathbf{w}}{{\mathbf{w}}^H}{\mathbf{H}}_{r,o}^H\mathbf{r}^H} \right\|} + {\sigma _r^2} \nonumber \\ &\ \ \quad \  = {{\mathbf{w}}^H}{\left( {\mathbf{H}_{r,o}^H}{\mathbf{r}^H}P_{B}\mathbf{r} {\mathbf{H}}_{r,o}\right)}{\mathbf{w}} + {\sigma _r^2},  \tag{\ref{P3_OF}{a}} \label{w_st1}  \\
& \ \ \xi_w \geq {\log _2 \left(1 + \frac{B}{\varrho_w} \right)}, \tag{\ref{P3_OF}{b}}  \\
& \ \ \varrho_w \leq  {P_{B}\left\| {{{\mathbf{h}}_{e,o}}{\mathbf{w}}{{\mathbf{w}}^H}{\mathbf{h}}_{e,o}^H} \right\|} + {\sigma _e^2} \nonumber \\ &\ \ \quad \ = {{\mathbf{w}}^H} \left( {{\mathbf{h}}_{e,o}^H}P_{B}{\mathbf{h}}_{e,o} \right) {\mathbf{w}} + {\sigma _e^2}. \tag{\ref{P3_OF}{c}} \label{w_st}
\end{align}
\end{small}

Although the objective function of \eqref{P3_OF} is concave with respect to $\xi_w$, it is still non-concave with respect to $\tau_w$, thus has to be convexified with its first order Taylor expansion as 
\begin{equation}
{\log _2 \left(1 + \frac{A}{\tau_w} \right)} \geq \log _2 P_{w} + \frac{Q_{w}}{P_{w}\ln{2}}\left (\tau_w - \overbar{\tau}_{w} \right), 
\label{P3_OF_v}
\end{equation}

\noindent where $P_{w} = {1 + A/\overbar{\tau}_{w}}$, $Q_{w} = - A/\overbar{\tau}_{w}^2$ and $\overbar{\tau}_{w}$ is the value of $\tau_w$ obtained in the last iteration.

In addition, we also convexify the constraint \eqref{w_st} with following approximation as 
\begin{small}
\begin{align}
    {{\mathbf{w}}^H} \left( {{\mathbf{h}}_{e,o}^H}P_{B}{\mathbf{h}}_{e,o} \right) {\mathbf{w}} \nonumber &\geq 2\text{Re}\left\{{{\mathbf{w}}^H}\left( {{\mathbf{h}}_{e,o}^H}P_{B}{\mathbf{h}}_{e,o} \right) {\mathbf{\overbar{w}}} \right\} \\ &\quad - {{\mathbf{\overbar{w}}}^H}\left( {{\mathbf{h}}_{e,o}^H}P_{B}{\mathbf{h}}_{e,o} \right) {\mathbf{\overbar{w}}}, \label{2Re}
\end{align}
\end{small}

\noindent where ${\mathbf{\overbar{w}}}$ denotes the value of $\mathbf{w}$ in the last iteration. With above operations, we can rewrite \eqref{P2_OF} as follows:  
\begin{small}
\begin{align}
\mathop {\max }\limits_{\mathbf{w},\xi_w, \tau_w, \varrho_w}&\ \ \log _2 P_{w} + \frac{Q_{w}}{P_{w}\ln{2}}\left (\tau_w - \overbar{\tau}_{w} \right) - \xi_w    \label{P_w} \\
{\text{s.t.}}\ \ & \ \   \tau_w \geq  {{\mathbf{w}}^H}{\left( {\mathbf{H}_{r,o}^H}{\mathbf{r}^H}P_{B}\mathbf{r} {\mathbf{H}}_{r,o}\right)}{\mathbf{w}} + {\sigma _r^2}, \tag{\ref{P_w}{a}} \\
& \ \ \xi_w \geq {\log _2 \left(1 + \frac{B}{\varrho_w} \right)}, \tag{\ref{P_w}{b}} \\
& \ \ \varrho_w \leq  2\text{Re}\left\{{{\mathbf{w}}^H} \left( {{\mathbf{h}}_{e,o}^H}P_{B}{\mathbf{h}}_{e,o} \right) {\mathbf{\overbar{w}}} \right\} \nonumber \\ & \ \ \quad \quad \ - {{\mathbf{\overbar{w}}}^H} \left( {{\mathbf{h}}_{e,o}^H}P_{B}{\mathbf{h}}_{e,o} \right) {\mathbf{\overbar{w}}} + {\sigma _e^2}, \tag{\ref{P_w}{c}} \label{w_st_r} \\
& \ \ {\text{Tr}}\left( {{\mathbf{w}}{{\mathbf{w}}^H}} \right) \leq 1, \tag{\ref{P_w}{d}} \\
& \ \ P_{B}{{\mathbf{w}}^H}{\left( {\mathbf{H}_{r,o}^H}{\mathbf{H}}_{r,o}\right)}{\mathbf{w}} \leq P_{th}, \tag{\ref{P_w}{e}}
\end{align}
\end{small}

\noindent which is a convex problem with convex constraints thus can be solved by using convex optimization tools such as CVX. 

\subsection{Optimizing receive beamforming vector $\mathbf{r}$}

With given $\mathbf{R}_o$, $\mathbf{T}_o$ and $\mathbf{w}$, we rewrite \eqref{P1_OF} for the optimization of $\mathbf{r}$ as follows: 
\begin{small}
\begin{align}
\mathop {\max }\limits_{\mathbf{r}}&\ \ {\log _2 \left(1 + \frac{{{P_{A}\mathbf{r}\mathbf{h}}_{ar}^H}{\mathbf{h}}_{ar}}{P_{B}{\left\| {\mathbf{r}{{\mathbf{H}}_{r,o}}{\mathbf{w}}{{\mathbf{w}}^H}{\mathbf{H}}_{r,o}^H\mathbf{r}^H} \right\|} + {\sigma _r^2}} \right)} 
- R_{e,o}  \label{P3_ori} \\
{\text{s.t.}}&\ \  \ \  {\text{Tr}}\left( {{\mathbf{r}}{{\mathbf{r}}^H}} \right) \leq 1. \tag{\ref{P3_ori}{a}} 
\end{align}
\end{small}

To tackle the non-concave objective function in \eqref{P3_ori}, we introduce slack variables $\psi_r$ and $\tau_r$ to transform the objective function and relative constraints as
\begin{small}
\begin{align}
\mathop {\max }\limits_{\mathbf{r},\psi_r, \tau_r}&\ \ {\log _2 \left(1 + \frac{\psi_r}{\tau_r} \right)} 
- R_{e,o}  \label{P3_OF2} \\
{\text{s.t.}}\ \  & \ \  \psi_r \leq  {{{P_{A}\mathbf{r}\mathbf{h}}_{ar}^H}{\mathbf{h}}_{ar}}\mathbf{r}^H, \tag{\ref{P3_OF2}{a}} \label{w_st2}  \\
& \ \ \tau_r \, \geq {P_{B}\left\| {\mathbf{r}{{\mathbf{H}}_{r,o}}{\mathbf{w}}{{\mathbf{w}}^H}{\mathbf{H}}_{r,o}^H\mathbf{r}^H} \right\|} + {\sigma _r^2} \nonumber \\ &\ \ \quad = {{\mathbf{w}}^H}{\left( {\mathbf{H}_{r,o}^H}{\mathbf{r}^H}P_{B}\mathbf{r} {\mathbf{H}}_{r,o}\right)}{\mathbf{w}} + {\sigma _r^2}. \tag{\ref{P3_OF2}{b}}
\end{align}
\end{small}

Since the objective function of \eqref{P3_OF2} is still non-concave with respect to $\tau_r$, we rewrite the first term of the objective function with a lower bound value by applying the inequality proved in (72) of \cite{Convex_ineq} for $x = 1/\psi_r$, $y = \tau_r$, $\overbar{x} = 1/\overbar{\psi}_{r}$ and $\overbar{y} = 1/\overbar{\tau}_{r}$ as
\begin{small}
\begin{align}
   \ln \left(1 + \frac{\psi_r}{\tau_r} \right) &=  \ln \left(1 + \frac{1}{xy} \right) \nonumber \\ 
   &\geq \ln \left(1 + \frac{1}{\overbar{x}\overbar{y}} \right) + \frac{1/ \overbar{x} \overbar{y}}{1+1/\overbar{x}\overbar{y}} \left(2 - \frac{x}{\overbar{x}} - \frac{y}{\overbar{y}} \right) \nonumber \\ 
   &=\ln \left(1 + \frac{\overbar{\psi}_{r}}{\overbar{\tau}_{r}} \right) + \frac{\overbar{\psi}_{r}/\overbar{\tau}_{r}}{1+\overbar{\psi}_{r}/\overbar{\tau}_{r}} \left(2 - \frac{\overbar{\psi}_{r}}{\psi_r} - \frac{\tau_r}{\overbar{\tau}_{r}} \right). \label{lb_P3}
\end{align}
\end{small}

The lower bounded function in \eqref{lb_P3} is then concave with respect to both $\psi_r$ and $\tau_r$, where $\overbar{\psi}_{r}$ and $\overbar{\tau}_{r}$ are the values of $\psi_r$ and $\tau_r$ obtained in the last iteration. 

However, since the \eqref{w_st2} is still non-convex, we use the first order Taylor expansions to express it as 
\begin{small}
\begin{align}
    {{\mathbf{r}\mathbf{h}}_{ar}^H}{\mathbf{h}}_{ar}\mathbf{r}^H \geq 2\text{Re}\{{{\mathbf{r}}} \left( \mathbf{h}_{ar}^H{\mathbf{h}}_{ar} \right) {\mathbf{\overbar{r}}^H} \} - {{{\mathbf{\overbar{r}}}} \left( \mathbf{h}_{ar}^H{\mathbf{h}}_{ar} \right) {\mathbf{\overbar{r}}^H} }
\end{align}
\end{small}

\noindent where ${\mathbf{\overbar{r}}}$ represents the values of $\mathbf{r}$ in the last iteration. Then, we can rewrite \eqref{P3_OF2} as follows:  
\begin{small}
\begin{align}
\mathop {\max }\limits_{\mathbf{r},\psi_r, \tau_r}&\ \ \log_2 \left(1 + \frac{\overbar{\psi}_{r}}{\overbar{\tau}_{r}} \right) + \frac{\frac{\overbar{\psi}_{r}}{\overbar{\tau}_{r}}}{\ln{2} (1+\frac{\overbar{\psi}_{r}}{\overbar{\tau}_{r}})}\left(2 - \frac{\overbar{\psi}_{r}}{\psi_r} - \frac{\tau_r}{\overbar{\tau}_{r}} \right) 
- R_{e,o}    \label{P_r} \\
{\text{s.t.}}\ \ & \ \   \psi_r \leq P_{A}\left(2\text{Re} \left \{{{\mathbf{r}}} \left( \mathbf{h}_{ar}^H{\mathbf{h}}_{ar} \right) {\mathbf{\overbar{r}}^H} \right\} - {{{\mathbf{\overbar{r}}}} \left( \mathbf{h}_{ar}^H{\mathbf{h}}_{ar} \right) {\mathbf{\overbar{r}}^H} } \right), \tag{\ref{P_r}{a}} \\
& \ \ \tau_r \, \geq  {{\mathbf{w}}^H}{\left( {\mathbf{H}_{r,o}^H}{\mathbf{r}^H}P_{B}\mathbf{r} {\mathbf{H}}_{r,o}\right)}{\mathbf{w}} + {\sigma _r^2}, \tag{\ref{P_r}{b}} \\
& \ \ {\text{Tr}}\left( {{\mathbf{r}}{{\mathbf{r}}^H}} \right) \leq 1, \tag{\ref{P_r}{c}} 
\end{align}
\end{small}

\noindent which is a convex problem with convex constraints can thus be solved efficiently by using CVX.

\subsection{Optimizing reflecting and refracting coefficient matrices $\mathbf{R}_o$ and $\mathbf{T}_o$}
In this subsection, we derive the coefficient matrices for both the ES and MS modes. Amplitudes and phase shifts matrices $\mathbf{R}_s$ and $\mathbf{T}_s$ will be calculated for the ES mode, while phase shifts matrices $\mathbf{R}_m'$, $\mathbf{T}_m'$ and mode selection matrix $\mathbf{A}_m$ will be calculated for the MS mode. 

\subsubsection{Optimizing $\mathbf{R}_s$ and $\mathbf{T}_s$ for ES-RIS}\hfill

With given $\mathbf{w}$ and $\mathbf{r}$, the secrecy capacity maximization with the ES-RIS in \eqref{P1_OF} can be rewritten as 
\begin{small}\begin{align}
\mathop {\max }\limits_{\mathbf{R}_s,\mathbf{T}_s}&\ \ {\log _2 \left(1 + \frac{A}{{P_{B}\left\| {\mathbf{r}{{\mathbf{H}}_{r,s}}{\mathbf{w}}{{\mathbf{w}}^H}{\mathbf{H}}_{r,s}^H\mathbf{r}^H} \right\|} + {\sigma _r^2}} \right)} \nonumber \\
&\ - {\log _2 \left(1 + \frac{B}{{P_{B}\left\| {{{\mathbf{h}}_{e,s}}{\mathbf{w}}{{\mathbf{w}}^H}{\mathbf{h}}_{e,s}^H} \right\|} + {\sigma _e^2}} \right)}  \label{P4_OF}\\
{\text{s.t.}}\ \ & \ \   u_{s,l}^2 + v_{s,l}^2 \leq 1,  \tag{\ref{P4_OF}{a}}  \label{P4_OF1} \\
& \ \  0 \leq{u_{s,l}},{v_{s,l}} \leq 1,
\tag{\ref{P4_OF}{b}}  \label{P4_OF2} \\
& \ \  0 \leq {\mu _{s,l}},{\nu _{s,l}} \leq 2\pi, \tag{\ref{P4_OF}{c}}  \\
& \ \ P_{B}{{\mathbf{w}}^H}{\left( {\mathbf{H}_{r,s}^H}{\mathbf{H}}_{r,s}\right)}{\mathbf{w}} \leq P_{th}. \tag{\ref{P4_OF}{d}} \label{P4_OF3}
\end{align}\end{small}

Since the constraints (\ref{P4_OF}{a}) - (\ref{P4_OF}{c}) are not convex, we introduce new variables $\mathbf{\mu}_{s} = \left\{ {{u_{s,1}}{e^{j{\mu _{s,1}}}},{u_{s,2}}{e^{j{\mu _{s,2}}}}, \ldots ,{u_{s,L}}{e^{j{\mu _{s,L}}}}} \right\} \in {\mathbb{C}^{1 \times L}}$ and $\mathbf{\nu}_{s} = \left\{ {{v_{s,1}}{e^{j{\nu _{s,1}}}},{v_{s,2}}{e^{j{\nu _{s,2}}}}, \ldots ,{v_{s,L}}{e^{j{\nu _{s,L}}}}} \right\}\in{\mathbb{C}^{1 \times L}}$ to transform those constraints to an equivalent one as 
\begin{equation}
\text{diag}({\mathbf{\mu}_{s}\mathbf{\mu}_{s}^H}+\mathbf{\nu}_{s}\mathbf{\nu}_{s}^H) \leq  \mathbf{1}_{L}, \label{P4_con}
\end{equation}

\noindent which is convex with respect to both $\mathbf{\mu}_{s}$ and $\mathbf{\nu}_{s}$.

Then we represent the objective function in \eqref{P4_OF} with $\mathbf{\mu}_{s}$ and $\mathbf{\nu}_{s}$. Firstly, we extend ${\left\| {\mathbf{r}{{\mathbf{H}}_{r,s}}{\mathbf{w}}{{\mathbf{w}}^H}{\mathbf{H}}_{r,s}^H\mathbf{r}^H} \right\|}$ in \eqref{P4_OF} as follows:
\begin{small}\begin{align}
&\left\| {\mathbf{r}{{\mathbf{H}}_{r,s}}{\mathbf{w}}{{\mathbf{w}}^H}{\mathbf{H}}_{r,s}^H\mathbf{r}^H} \right\| = {{\mathbf{w}}^H}{\left( {\mathbf{H}_{r,s}^H}{\mathbf{r}^H}\mathbf{r} {\mathbf{H}}_{r,s}\right)}{\mathbf{w}} \nonumber \\
&= {{\mathbf{w}}^H}{\left( \mathbf{H}_{ti}^H\mathbf{R}_{s}^H\mathbf{H}_{ir} + \mathbf{H}_{tr}^H\right){\mathbf{r}^H}\mathbf{r} \left(\mathbf{H}_{tr} + \mathbf{H}_{ir}^H\mathbf{R}_{s}\mathbf{H}_{ti}\right)}{\mathbf{w}} \nonumber \\
&= {\mathbf{w}^H\mathbf{H}_{ti}^H\mathbf{R}_{s}^H\mathbf{H}_{ir}{\mathbf{r}^H}\mathbf{r}\mathbf{H}_{ir}^H\mathbf{R}_{s}\mathbf{H}_{ti}\mathbf{w}} + {\mathbf{w}^H\mathbf{H}_{tr}^H{\mathbf{r}^H}\mathbf{r}\mathbf{H}_{tr}\mathbf{w}} \nonumber \\
&\quad \ \!+\! {\mathbf{w}^H\mathbf{H}_{ti}^H\mathbf{R}_{s}^H\mathbf{H}_{ir}{\mathbf{r}^H}\mathbf{r}\mathbf{H}_{tr}\mathbf{w}} \!+\! {\mathbf{w}^H\mathbf{H}_{tr}^H{\mathbf{r}^H}\mathbf{r}\mathbf{H}_{ir}^H\mathbf{R}_{s}\mathbf{H}_{ti}\mathbf{w}}.
\label{P4_1}
\end{align}\end{small}

Since all the four extended terms in \eqref{P4_1} are scalars which equal to their traces, the above expression can be equivalently expressed as a summation of trace as 
\begin{small}\begin{align}
&\text{Tr}\left({\mathbf{w}^H\mathbf{H}_{ti}^H\mathbf{R}_{s}^H\mathbf{H}_{ir}{\mathbf{r}^H}\mathbf{r}\mathbf{H}_{ir}^H\mathbf{R}_{s}\mathbf{H}_{ti}\mathbf{w}}\right) \!+\! \text{Tr}\left({\mathbf{w}^H\mathbf{H}_{tr}^H{\mathbf{r}^H}\mathbf{r}\mathbf{H}_{tr}\mathbf{w}}\right) \nonumber \\
&+\! \text{Tr}\left({\mathbf{w}^H\mathbf{H}_{ti}^H\mathbf{R}_{s}^H\mathbf{H}_{ir}{\mathbf{r}^H}\mathbf{r}\mathbf{H}_{tr}\mathbf{w}}\right) \!+\! \text{Tr}\left({\mathbf{w}^H\mathbf{H}_{tr}^H{\mathbf{r}^H}\mathbf{r}\mathbf{H}_{ir}^H\mathbf{R}_{s}\mathbf{H}_{ti}\mathbf{w}}\right) \nonumber \\
&= \text{Tr}\left(\mathbf{R}_{s}^H\mathbf{X}_{s,1}\mathbf{R}_{s}\mathbf{Y}_{s,1}\right) \!+\! \text{Tr}\left(\mathbf{R}_{s}\mathbf{Z}_{s,1}\right) \!+\! \text{Tr}\left(\mathbf{R}_{s}^H\mathbf{Z}_{s,1}^H\right) \!+\! q_{s,1},
\label{P4_2}
\end{align}\end{small}

\noindent where $\mathbf{X}_{s,1} = \mathbf{H}_{ir}{\mathbf{r}^H}\mathbf{r}\mathbf{H}_{ir}^H$, $\mathbf{Y}_{s,1} = \mathbf{H}_{ti}\mathbf{w}\mathbf{w}^H\mathbf{H}_{ti}^H$, $\mathbf{Z}_{s,1} = \mathbf{H}_{ti}\mathbf{w}\mathbf{w}^H\mathbf{H}_{tr}^H{\mathbf{r}^H}\mathbf{r}\mathbf{H}_{ir}^H$ and $q_{s,1} = \mathbf{w}^H\mathbf{H}_{tr}^H{\mathbf{r}^H}\mathbf{r}\mathbf{H}_{tr}\mathbf{w}$, while the cyclicality of trace is applied to the transformation in \eqref{P4_2}.

Based on the matrix properties in \cite[Eq. (1.10.6)]{Matrix_Analysis}, we transform \eqref{P4_2} into a function of $\mathbf{\mu}_{s}$ as 
\begin{small}\begin{align}
&\text{Tr}\left(\mathbf{R}_{s}^H\mathbf{X}_{s,1}\mathbf{R}_{s}\mathbf{Y}_{s,1}\right) + \text{Tr}\left(\mathbf{R}_{s}\mathbf{Z}_{s,1}\right) + \text{Tr}\left(\mathbf{R}_{s}^H\mathbf{Z}_{s,1}^H\right) + q_{s,1} \nonumber \\
&= \mathbf{\mu}_{s}^H \left(\mathbf{X}_{s,1}\odot\mathbf{Y}_{s,1}\right)\mathbf{\mu}_{s} + \text{diag}\{\mathbf{Z}_{s,1}^H\}{\mu}_{s}^* + \mathbf{\mu}_{s}^T\text{diag}\{\mathbf{Z}_{s,1}\} + q_{s,1} \nonumber \\
&= \mathbf{\mu}_{s}^H \left(\mathbf{X}_{s,1}\odot\mathbf{Y}_{s,1}\right)\mathbf{\mu}_{s} +   2\text{Re}\left \{ \mathbf{\mu}_{s}^T\text{diag}\{\mathbf{Z}_{s,1}\} \right \} + q_{s,1}.
\label{P4_mu}
\end{align}\end{small}

Similarly, we can transform $\left\| {{{\mathbf{h}}_{e,s}}{\mathbf{w}}{{\mathbf{w}}^H}{\mathbf{h}}_{e,s}^H} \right\|$ in \eqref{P4_OF} as
\begin{small}\begin{align}
&\left\| {{{\mathbf{h}}_{e,s}}{\mathbf{w}}{{\mathbf{w}}^H}{\mathbf{h}}_{e,s}^H} \right\| = {{\mathbf{w}}^H} \left( {{\mathbf{h}}_{e,s}^H}{\mathbf{h}}_{e,s} \right) {\mathbf{w}} \nonumber \\
&= \text{Tr}({{\mathbf{w}}^H} \left( {{\mathbf{h}}_{e,s}^H}{\mathbf{h}}_{e,s} \right) {\mathbf{w}}) = \text{Tr}({{\mathbf{w}}^H}{\mathbf{H}}_{ti}^H{\mathbf{T}}_s^H{\mathbf{h}}_{ie}{{\mathbf{h}}_{ie}^H}{\mathbf{T}}_s{\mathbf{H}}_{ti}{\mathbf{w}}) \nonumber \\
&= \text{Tr}({\mathbf{T}}_s^H{\mathbf{X}}_{s,2}{\mathbf{T}}_s{\mathbf{Y}}_{s,2}) = \mathbf{\nu}_{s}^H \left(\mathbf{X}_{s,2}\odot\mathbf{Y}_{s,2}\right)\mathbf{\nu}_{s},
\label{P4_nu}
\end{align}\end{small}

\noindent where $\mathbf{X}_{s,2} = {\mathbf{h}}_{ie}{{\mathbf{h}}_{ie}^H}$ and $\mathbf{Y}_{s,2} = {\mathbf{H}}_{ti}{\mathbf{w}}{{\mathbf{w}}^H}{\mathbf{H}}_{ti}^H$.

In addition, we transform ${{\mathbf{w}}^H}{\left( {\mathbf{H}_{r,s}^H}{\mathbf{H}}_{r,s}\right)}{\mathbf{w}}$ in \eqref{P4_OF3} by using the expressions of \eqref{P4_1} - \eqref{P4_mu} as 

\begin{small}\begin{align}
{{\mathbf{w}}^H}{\left( {\mathbf{H}_{r,s}^H}{\mathbf{H}}_{r,s}\right)}{\mathbf{w}} &= \mathbf{\mu}_{s}^H \left(\mathbf{X}_{s,3}\odot\mathbf{Y}_{s,3}\right)\mathbf{\mu}_{s} \nonumber \\ & \quad + 2\text{Re}\left \{\mathbf{\mu}_{s}^T\text{diag}\{\mathbf{Z}_{s,3}\}\right \} + q_{s,3},
\label{P4_d}
\end{align}\end{small}

\noindent where $\mathbf{X}_{s,3} = \mathbf{H}_{ir}\mathbf{H}_{ir}^H$, $\mathbf{Y}_{s,3} = \mathbf{H}_{ti}\mathbf{w}\mathbf{w}^H\mathbf{H}_{ti}^H$, $\mathbf{Z}_{s,3} = \mathbf{H}_{ti}\mathbf{w}\mathbf{w}^H\mathbf{H}_{tr}^H\mathbf{H}_{ir}^H$ and $q_{s,3} = \mathbf{w}^H\mathbf{H}_{tr}^H\mathbf{H}_{tr}\mathbf{w}$.

However, even with the transformation of \eqref{P4_mu} and \eqref{P4_nu}, the objective function of \eqref{P4_OF} is still non-concave with respect to $\mathbf{\mu}_{s}$ and $\mathbf{\nu}_{s}$. We thus have to introduce slack variables to rewrite \eqref{P4_OF} as  
\begin{small}\begin{align}
\mathop {\max }\limits_{\mathbf{\mu}_{s}, \mathbf{\nu}_{s}, \tau_s, \xi_s, \varrho_s}&\ \ {\log _2 \left(1 + \frac{A}{\tau_{s}} \right)} - \xi_{s} \label{P4_r} \\
{\text{s.t.}}\ \ & \ \   \tau_s \geq P_B \left. (\mathbf{\mu}_{s}^H \left(\mathbf{X}_{s,1}\odot\mathbf{Y}_{s,1}\right)\mathbf{\mu}_{s} \right. \nonumber \\
&\ \ \quad \quad \left. + 2\text{Re} \{\mathbf{\mu}_{s}^T\text{diag}\{\mathbf{Z}_{s,1}\} \} + q_{s,1}\right.) + {\sigma _r^2}, \tag{\ref{P4_r}{a}} \label{P4_r_a}\\
& \ \ \xi_s \geq {\log _2 \left(1 + \frac{B}{\varrho_s} \right)}, \tag{\ref{P4_r}{b}} \label{P4_r_b}\\
& \ \ \varrho_s \leq P_B\mathbf{\nu}_{s}^H \left(\mathbf{X}_{s,2}\odot\mathbf{Y}_{s,2}\right)\mathbf{\nu}_{s} + {\sigma _e^2}, \tag{\ref{P4_r}{c}} \label{P4_r_c} \\
& \ \ \text{diag}({\mathbf{\mu}_{s}\mathbf{\mu}_{s}^H}+\mathbf{\nu}_{s}\mathbf{\nu}_{s}^H) \leq  \mathbf{1}_{L},  \tag{\ref{P4_r}{d}} \label{P4_r_d} \\
& \ \ P_B \left. (\mathbf{\mu}_{s}^H \left(\mathbf{X}_{s,3}\odot\mathbf{Y}_{s,3}\right)\mathbf{\mu}_{s} \right. \nonumber \\
&\ \ \left. + 2\text{Re} \{\mathbf{\mu}_{s}^T\text{diag}\{\mathbf{Z}_{s,3}\}\} + q_{s,3}\right.) \leq P_{th}. \tag{\ref{P4_r}{e}} \label{P4_r_s}
\end{align}\end{small}

In addition, the first term in \eqref{P4_r} can be reformed to a concave function with respect to $\tau_s$ as follow 
\begin{small}\begin{align}
\log _2 \left(1 + \frac{\mathbf{A}}{\tau_{s}} \right) \geq \log _2 \left(P_{s}\right) + \frac{Q_{s}}{P_{s}\ln{2}} \left(\tau_{s} - \overbar{\tau}_{s} \right),
\end{align}\end{small}

\noindent where $P_{s} = 1 + \mathbf{A}/\overbar{\tau}_{s}$, $Q_{s} = - A/\overbar{\tau}_{s}^2$ and $\overbar{\tau}_{s}$ is the value obtained in the last iteration.

Also we transform the constraint of \eqref{P4_r_c} with $\overbar{\mathbf{\nu}}_s$ of the last iteration by using the method in \eqref{2Re}, to reform the whole problem in \eqref{P4_r} as follows:
\begin{small}\begin{align}
\mathop {\max }\limits_{\mathbf{\mu}_{s}, \mathbf{\nu}_{s}, \tau_s, \xi_s, \varrho_s}&\ \ \log _2 \left(P_{s}\right) + \frac{Q_{s}}{P_{s}\ln{2}} \left(\tau_{s} - \overbar{\tau}_{s} \right) - \xi_{s} \label{P_ES} \\
{\text{s.t.}}\ \ & \ \  \eqref{P4_r_a} - \eqref{P4_r_b},  \tag{\ref{P_ES}{a}} \\
& \ \ \varrho_s \leq 2\text{Re}\left\{P_B\mathbf{\nu}_{s}^H \left(\mathbf{X}_{s,2}\odot\mathbf{Y}_{s,2}\right)\overbar{\mathbf{\nu}}_{s} \right\} \nonumber \\ 
& \ \ \ \ \ \ \ \ - P_B\overbar{\mathbf{\nu}}_{s}^H \left(\mathbf{X}_{s,2}\odot\mathbf{Y}_{s,2}\right)\overbar{\mathbf{\nu}}_{s} + {\sigma _e^2}, \tag{\ref{P_ES}{b}} \\
& \ \ \eqref{P4_r_d} - \eqref{P4_r_s}, \tag{\ref{P_ES}{c}}
\end{align}\end{small}

\noindent which is a convex problem and is readily to be solved by using CVX. 

\subsubsection{Optimizing $\mathbf{R}_m$ and $\mathbf{T}_m$ for MS-RIS}\hfill

With given $\mathbf{w}$ and $\mathbf{r}$, the secrecy capacity maximization with the MS-RIS can be transformed as 
\begin{small}\begin{align}
\mathop {\max }\limits_{\mathbf{R}_m,\mathbf{T}_m}&\ \ {\log _2 \left(1 + \frac{A}{{P_{B}\left\| {\mathbf{r}{{\mathbf{H}}_{r,m}}{\mathbf{w}}{{\mathbf{w}}^H}{\mathbf{H}}_{r,m}^H\mathbf{r}^H} \right\|} + {\sigma _r^2}} \right)} \nonumber \\
&\ - {\log _2 \left(1 + \frac{B}{{P_{B}\left\| {{{\mathbf{h}}_{e,m}}{\mathbf{w}}{{\mathbf{w}}^H}{\mathbf{h}}_{e,m}^H} \right\|} + {\sigma _e^2}} \right)}  \label{P5_OF}\\
{\text{s.t.}}\ \ & \ \   u_{m,l} + v_{m,l} = 1,  \tag{\ref{P5_OF}{a}}  \label{P5_OF1} \\
& \ \  0 \leq{u_{m,l}},{v_{m,l}} \in \left\{0,1\right\},
\tag{\ref{P5_OF}{b}}  \label{P5_OF2} \\
& \ \  0 \leq {\mu _{m,l}},{\nu _{m,l}} \leq 2\pi, \tag{\ref{P5_OF}{c}}  
\label{P5_OF3} \\
& \ \ P_{B}{{\mathbf{w}}^H}{\left( {\mathbf{H}_{r,m}^H}{\mathbf{H}}_{r,m}\right)}{\mathbf{w}} \leq P_{th}. \tag{\ref{P5_OF}{d}}  
\label{P5_OF4}
\end{align}\end{small}

However, the objective function is non-concave since both $\mathbf{R}_m$ and $\mathbf{T}_m$ contain binary variables. To tackle this problem, we reform the above two matrices as $\mathbf{R}_m = {\mathbf{A}_m}{\mathbf{R}_m^{'}}$ and $\mathbf{T}_m = {\mathbf{B}_m}{\mathbf{T}_m^{'}}$. Here $\mathbf{A}_m = \text{diag}\left\{ {{u_{m,1}},{u_{m,2}}, \ldots ,{u_{m,L}}} \right\} \in {\mathbb{R}^{L \times L}}$ and $\mathbf{B}_m = \text{diag}\left\{ {{v_{m,1}},{v_{m,2}}, \ldots ,{v_{m,L}}} \right\} \in {\mathbb{R}^{L \times L}}$ are the amplitudes which mean the mode selection of reflecting or refracting, while ${\mathbf{R}_m^{'}} = \text{diag}\left\{ {{e^{j{\mu _{m,1}}}},{e^{j{\mu _{m,2}}}}, \ldots ,{e^{j{\mu _{m,L}}}}} \right\} \in {\mathbb{C}^{L \times L}} $ and ${\mathbf{T}_m^{'}} = \text{diag}\left\{ {{e^{j{\nu _{m,1}}}},{e^{j{\nu _{m,2}}}}, \ldots ,{e^{j{\nu _{m,L}}}}} \right\} \in {\mathbb{C}^{L \times L}} $ are the phase shifts for reflecting and refracting, respectively. 

With above reformation, we then divide \eqref{P5_OF} into two sub-problems by optimizing phase shifts and mode selection sequentially. 

Firstly, we optimize ${\mathbf{R}_m^{'}}$ and ${\mathbf{T}_m^{'}}$ with fixed $\mathbf{A}_m$ and $\mathbf{B}_m$. With the constraint in \eqref{P5_OF1}, it is obvious that $\mathbf{B}_m = \mathbf{I} - \mathbf{A}_m$. Therefore, the channel ${{\mathbf{H}}_{r,m}}$ and ${{\mathbf{h}}_{e,m}}$ can be rewritten as
\begin{small}\begin{align}
&{{\mathbf{H}}_{r,m}} = \mathbf{H}_{tr} + \mathbf{H}_{ir}^H\mathbf{A}_{m}\mathbf{R}_{m}^{'}\mathbf{H}_{ti} \\
&{{\mathbf{h}}_{e,m}} = {\mathbf{h}}_{ie}^H(\mathbf{I} - \mathbf{A}_m){\mathbf{T}_{m}^{'}}{{\mathbf{H}}_{ti}}. 
\end{align}\end{small}

Then with the similar operation by introducing new variables and transforming with trace format, we can have 
\begin{small}\begin{align}
&{\left\| {\mathbf{r}{{\mathbf{H}}_{r,m}}{\mathbf{w}}{{\mathbf{w}}^H}{\mathbf{H}}_{r,m}^H\mathbf{r}^H} \right\|} \nonumber \\
&= \mathbf{\mu}_{m}^H \left(\mathbf{X}_{m,1}\odot\mathbf{Y}_{m,1}\right)\mathbf{\mu}_{m} 
 + 2\text{Re}\{\mathbf{\mu}_{m}^T\text{diag}\{\mathbf{Z}_{m,1}\}\} + q_{m,1},
\end{align}\end{small}
where $\mathbf{\mu}_{m} = \left\{ {{e^{j{\mu _{m,1}}}},{e^{j{\mu _{m,2}}}}, \ldots ,{e^{j{\mu _{m,L}}}}} \right\} \in {\mathbb{C}^{1 \times L}}$ and 
\begin{small}\begin{align}
&\mathbf{X}_{m,1} = \mathbf{A}_{m}^{H}\mathbf{H}_{ir}{\mathbf{r}^H}\mathbf{r}\mathbf{H}_{ir}^H\mathbf{A}_{m}, \nonumber \\
&\mathbf{Y}_{m,1} = \mathbf{H}_{ti}\mathbf{w}\mathbf{w}^H\mathbf{H}_{ti}^H , \nonumber \\
&\mathbf{Z}_{m,1} = \mathbf{H}_{ti}\mathbf{w}\mathbf{w}^H\mathbf{H}_{tr}^H{\mathbf{r}^H}\mathbf{r}\mathbf{H}_{ir}^H\mathbf{A}_{m}, \nonumber \\
&q_{m,1} = \mathbf{w}^H\mathbf{H}_{tr}^H{\mathbf{r}^H}\mathbf{r}\mathbf{H}_{tr}\mathbf{w}. \nonumber  
\end{align}\end{small}
In the same manner, we transform $\left\| {{{\mathbf{h}}_{e,m}}{\mathbf{w}}{{\mathbf{w}}^H}{\mathbf{h}}_{e,m}^H} \right\|$ as 
\begin{small}\begin{align}
&\left\| {{{\mathbf{h}}_{e,m}}{\mathbf{w}}{{\mathbf{w}}^H}{\mathbf{h}}_{e,m}^H} \right\| = \mathbf{\nu}_{m}^H \left(\mathbf{X}_{m,2}\odot\mathbf{Y}_{m,2}\right)\mathbf{\nu}_{m},
\end{align}\end{small}
where $\mathbf{\nu}_{m} = \left\{ {{e^{j{\nu _{m,1}}}},{e^{j{\nu _{m,2}}}}, \ldots ,{e^{j{\nu _{m,L}}}}} \right\}\in{\mathbb{C}^{1 \times L}}$ and 
\begin{small}\begin{align}
&\mathbf{X}_{m,2} = (\mathbf{I} - \mathbf{A}_m)^H{\mathbf{h}}_{ie}{{\mathbf{h}}_{ie}^H}(\mathbf{I} - \mathbf{A}_m), \nonumber \\
&\mathbf{Y}_{m,2} = {\mathbf{H}}_{ti}{\mathbf{w}}{{\mathbf{w}}^H}{\mathbf{H}}_{ti}^H. \nonumber
\end{align}\end{small}

Also, ${{\mathbf{w}}^H}{\left( {\mathbf{H}_{r,m}^H}{\mathbf{H}}_{r,m}\right)}{\mathbf{w}}$ in \eqref{P5_OF4} can be transformed as 
\begin{small}\begin{align}
&{{\mathbf{w}}^H}{\left( {\mathbf{H}_{r,m}^H}{\mathbf{H}}_{r,m}\right)}{\mathbf{w}} \nonumber \\
&=\mathbf{\mu}_{m}^H \left(\mathbf{X}_{m,3}\odot\mathbf{Y}_{m,3}\right)\mathbf{\mu}_{m} + 2\text{Re}\{\mathbf{\mu}_{m}^T\text{diag}\{\mathbf{Z}_{m,3}\}\} + q_{m,3},
\end{align}\end{small}
where 
\begin{small}\begin{align}
\mathbf{X}_{m,3} &= \mathbf{A}_{m}^{H}\mathbf{H}_{ir}\mathbf{H}_{ir}^H\mathbf{A}_{m}, \nonumber \\
\mathbf{Y}_{m,3} &= \mathbf{H}_{ti}\mathbf{w}\mathbf{w}^H\mathbf{H}_{ti}^H , \nonumber \\
\mathbf{Z}_{m,3} &= \mathbf{H}_{ti}\mathbf{w}\mathbf{w}^H\mathbf{H}_{tr}^H\mathbf{H}_{ir}^H\mathbf{A}_{m}, \nonumber \\
q_{m,3} &= \mathbf{w}^H\mathbf{H}_{tr}^H\mathbf{H}_{tr}\mathbf{w}. \nonumber  
\end{align}\end{small}

After convexifing operations, the whole problem of $\mathbf{R}_{m}^{'}$ and $\mathbf{T}_{m}^{'}$ can be reformed as 
\begin{small}\begin{align}
\mathop {\max }\limits_{\mathbf{\mu}_{m}, \mathbf{\nu}_{m}, \tau_m, \xi_m, \varrho_m}&\ \ \log _2 \left(P_m\right) + \frac{Q_m}{P_{m}\ln{2}} \left(\tau_{m} - \overbar{\tau}_{m} \right) - \xi_{m} \label{P_MS} \\
{\text{s.t.}}\ \ & \ \   \tau_m \geq  P_B \left. (\mathbf{\mu}_{m}^H \left(\mathbf{X}_{m,1}\odot\mathbf{Y}_{m,1}\right)\mathbf{\mu}_{m}  \right. \nonumber \\
&\quad \quad \quad \left. + 2\text{Re}\{\mathbf{\mu}_{m}^T\text{diag}\{\mathbf{Z}_{m,1}\}\} + q_{m,1}\right.) + {\sigma _r^2}, \tag{\ref{P_MS}{a}} \\
& \ \ \xi_m \geq {\log _2 \left(1 + \frac{B}{\varrho_m} \right)}, \tag{\ref{P_MS}{b}} \\
& \ \ \varrho_m \leq 2\text{Re}\{P_B\mathbf{\nu}_{m}^H \left(\mathbf{X}_{m,2}\odot\mathbf{Y}_{m,2}\right)\overbar{\mathbf{\nu}}_{m} \} \nonumber \\ 
&\quad \quad \quad - P_B\overbar{\mathbf{\nu}}_{m}^H \left(\mathbf{X}_{m,2}\odot\mathbf{Y}_{m,2}\right)\overbar{\mathbf{\nu}}_{m} + {\sigma _e^2}, \tag{\ref{P_MS}{c}} \\
& \ \ \text{diag}({\mathbf{\mu}_{m}\mathbf{\mu}_{m}^H}) \leq  \mathbf{1}_{L}, \tag{\ref{P_MS}{d}} \\
& \ \ \text{diag}(\mathbf{\nu}_{m}\mathbf{\nu}_{m}^H) \leq  \mathbf{1}_{L}, \tag{\ref{P_MS}{e}} \\
& \ \ P_B \left. (\mathbf{\mu}_{m}^H \left(\mathbf{X}_{m,3}\odot\mathbf{Y}_{m,3}\right)\mathbf{\mu}_{m} \right. \nonumber \\
&\ \ \left. + 2\text{Re}\{\mathbf{\mu}_{m}^T\text{diag}\{\mathbf{Z}_{m,3}\}\} + q_{m,3}\right.) \leq  P_{th}, \tag{\ref{P_MS}{f}}
\end{align}\end{small}

\noindent where $P_{m} = 1 + A/\overbar{\tau}_{m}$, $Q_{m} = - A/\overbar{\tau}_{m}^2$ and $\overbar{\tau}_{m}$ is the value obtained in the last iteration. 

\subsubsection{Optimizing Mode selection matrix $\mathbf{A}_m$ for MS-RIS}\hfill

With given $\mathbf{w}$, $\mathbf{r}$, $\mathbf{R}_{m}^{'}$ and $\mathbf{T}_{m}^{'}$, the problem of optimizing $\mathbf{A}_m$ can then be formulated as 
\begin{small}\begin{align}
\mathop {\max }\limits_{\mathbf{A}_m}&\ \ {\log _2 \left(1 + \frac{A}{{P_{B}\left\| {\mathbf{r}{{\mathbf{H}}_{r,m}}{\mathbf{w}}{{\mathbf{w}}^H}{\mathbf{H}}_{r,m}^H\mathbf{r}^H} \right\|} + {\sigma _r^2}} \right)} \nonumber \\
&- {\log _2 \left(1 + \frac{B}{{P_{B}\left\| {{{\mathbf{h}}_{e,m}}{\mathbf{w}}{{\mathbf{w}}^H}{\mathbf{h}}_{e,m}^H} \right\|} + {\sigma _e^2}} \right)}  \label{P6}\\
{\text{s.t.}}\ \ & \ \   u_{m,l} + v_{m,l} = 1,  \tag{\ref{P6}{a}}  \label{P6_OF1} \\
& \ \  {u_{m,l}},{v_{m,l}} \in \left\{0,1\right\},
\tag{\ref{P6}{b}}  \label{P6_OF2} \\
& \ \  P_{B}{{\mathbf{w}}^H}{\left( {\mathbf{H}_{r,m}^H}{\mathbf{H}}_{r,m}\right)}{\mathbf{w}} \leq P_{th}, \tag{\ref{P6}{c}}  
\label{P6_OF3}
\end{align}\end{small}

To tackle the binary variables in the objective function, we firstly introduce a new vector ${{\mathbf{a}}_m} = \text{diag}\{\mathbf{A}_{m}\} \in {\mathbb{R}^{L \times 1}}$. Then ${\left\| {\mathbf{r}{{\mathbf{H}}_{r,m}}{\mathbf{w}}{{\mathbf{w}}^H}{\mathbf{H}}_{r,m}^H\mathbf{r}^H} \right\|}$ in the objective function of \eqref{P6} can be rewritten as 
\begin{small}\begin{align} 
&\left\| {\mathbf{r}{{\mathbf{H}}_{r,m}}{\mathbf{w}}{{\mathbf{w}}^H}{\mathbf{H}}_{r,m}^H\mathbf{r}^H} \right\|  \nonumber \\
&= {{\mathbf{w}}^H}{\left( \mathbf{H}_{ti}^H\mathbf{R}_{m}'^H\mathbf{A}_{m}^H\mathbf{H}_{ir} + \mathbf{H}_{tr}^H\right){\mathbf{r}^H}\mathbf{r} \left(\mathbf{H}_{tr} + \mathbf{H}_{ir}^H\mathbf{A}_{m}\mathbf{R}_{m}'\mathbf{H}_{ti}\right)}{\mathbf{w}} \nonumber \\
&= \text{Tr}\left(\mathbf{A}_{m}^H\mathbf{X}_{a,1}\mathbf{A}_{m}\mathbf{Y}_{a,1}\right) + \text{Tr}\left(\mathbf{A}_{m}\mathbf{Z}_{a,1}\right) + \text{Tr}\left(\mathbf{A}_{m}^H\mathbf{Z}_{a,1}^H\right) + q_{a,1} \nonumber \\
&= \mathbf{a}_{m}^H \Pi_{a,1}\mathbf{a}_{m} + \text{diag}\{\mathbf{Z}_{a,1}^H\}\mathbf{a}_{m}^* + \mathbf{a}_{m}^T\text{diag}\{\mathbf{Z}_{a,1}\} + q_{a,1} \nonumber \\
&= \mathbf{a}_{m}^H \Pi_{a,1}\mathbf{a}_{m} + 2\text{Re}\{\mathbf{a}_{m}^T\text{diag}\{\mathbf{Z}_{a,1}\}\}+ q_{a,1},
\label{SI_withA}
\end{align}\end{small}

\noindent where ${{\mathbf{X}}_{a,1}} = {{\mathbf{H}}_{ir}}{{\mathbf{r}}^H}\mathbf{r}{\mathbf{H}}_{ir}^H$, ${{\mathbf{Y}}_{a,1}} = {{{\mathbf{R}}}_m'}{{\mathbf{H}}_{ti}}{\mathbf{w}}{{\mathbf{w}}^H}{\mathbf{H}}_{ti}^H{{{\mathbf{R}}}_m'^H}$, ${{\mathbf{Z}}_{a,1}} = {{{\mathbf{R}}}_m'}{{\mathbf{H}}_{ti}}{\mathbf{w}}{{\mathbf{w}}^H}{{\mathbf{H}}_{tr}^H}{\mathbf{H}}_{ir}^H$,$q_{a,1} = \mathbf{w}^H\mathbf{H}_{tr}^H{\mathbf{r}^H}\mathbf{r}\mathbf{H}_{tr}\mathbf{w}$ and $\Pi_{a,1} = \mathbf{X}_{a,1}\odot\mathbf{Y}_{a,1}$. 

Similarly, $\left\| {{{\mathbf{h}}_{e,m}}{\mathbf{w}}{{\mathbf{w}}^H}{\mathbf{h}}_{e,m}^H} \right\|$ in the objective function can be transformed as 
\begin{small}\begin{align} 
&\left\| {{{\mathbf{h}}_{e,m}}{\mathbf{w}}{{\mathbf{w}}^H}{\mathbf{h}}_{e,m}^H} \right\| \nonumber \\
&= {{\mathbf{w}}^H}{\mathbf{H}}_{ti}^H{\mathbf{T}_m'^H}\left( {{\mathbf{I}} \!-\! {{\mathbf{A}}_m}}\right)^H {{\mathbf{h}}_{ie}}{\mathbf{h}}_{ie}^H\left( {{\mathbf{I}}\! - \!{{\mathbf{A}}_m}} \right){\mathbf{T}_m'}\!{{\mathbf{H}}_{ti}}{\mathbf{w}} \nonumber \\
&= \text{Tr}\left(\mathbf{A}_{m}^H\mathbf{X}_{a,2}\mathbf{A}_{m}\mathbf{Y}_{a,2}\right) - \text{Tr}\left(\mathbf{A}_{m}\mathbf{Z}_{a,2}\right) - \text{Tr}\left(\mathbf{A}_{m}^H\mathbf{Z}_{a,2}^H\right) + q_{a,2} \nonumber \\
&= \mathbf{a}_{m}^H\Pi_{a,2}\mathbf{a}_{m} - 2\text{Re}\{\mathbf{a}_{m}^T\text{diag}\{\mathbf{Z}_{a,2}\}\}+ q_{a,2},
\label{Jam_Power}
\end{align}\end{small}

\noindent where ${{\mathbf{X}}_{a,2}} = {{\mathbf{h}}_{ie}}{\mathbf{h}}_{ie}^H$, ${{\mathbf{Y}}_{a,2}} = {{{\mathbf{T}_m'}}}{{\mathbf{H}}_{ti}}{\mathbf{w}}{{\mathbf{w}}^H}{\mathbf{H}}_{ti}^H{{{\mathbf{T}}}_m'^H}$, ${{\mathbf{Z}}_{a,2}} = {{{\mathbf{T}}}_m'}{{\mathbf{H}}_{ti}}{\mathbf{w}}{{\mathbf{w}}^H}{{\mathbf{H}}_{ti}^H}{{\mathbf{T}}_m'^H}{{\mathbf{h}}_{ie}}{\mathbf{h}}_{ie}^H$, $\Pi_{a,2} = \mathbf{X}_{a,2}\odot\mathbf{Y}_{a,2}$ and $q_{a,2} = {\mathbf{h}}_{ie}^H{{{\mathbf{T}}}_m'}{{\mathbf{H}}_{ti}}{\mathbf{w}}{{\mathbf{w}}^H}{{\mathbf{H}}_{ti}^H}{{\mathbf{T}}_m'^H}{{\mathbf{h}}_{ie}}$.

Also, the ${{\mathbf{w}}^H}{\left( {\mathbf{H}_{r,m}^H}{\mathbf{H}}_{r,m}\right)}{\mathbf{w}}$ in \eqref{P6_OF3} has to be rewritten as 
\begin{small}\begin{align} 
&{{\mathbf{w}}^H}{\left( \mathbf{H}_{ti}^H\mathbf{R}_{m}'^H\mathbf{A}_{m}^H\mathbf{H}_{ir} + \mathbf{H}_{tr}^H\right)\left(\mathbf{H}_{tr} + \mathbf{H}_{ir}^H\mathbf{A}_{m}\mathbf{R}_{m}'\mathbf{H}_{ti}\right)}{\mathbf{w}} \nonumber \\
&= \text{Tr}\left(\mathbf{A}_{m}^H\mathbf{X}_{a,3}\mathbf{A}_{m}\mathbf{Y}_{a,3}\right) + \text{Tr}\left(\mathbf{A}_{m}\mathbf{Z}_{a,3}\right) + \text{Tr}\left(\mathbf{A}_{m}^H\mathbf{Z}_{a,3}^H\right) + q_{a,3} \nonumber \\
&= \mathbf{a}_{m}^H\Pi_{a,3}\mathbf{a}_{m} + 2\text{Re}\{\mathbf{a}_{m}^T\text{diag}\{\mathbf{Z}_{a,3}\}\}+ q_{a,3},
\label{rPower_limit}
\end{align}\end{small}where ${{\mathbf{X}}_{a,3}} = {{\mathbf{H}}_{ir}}{\mathbf{H}}_{ir}^H$, ${{\mathbf{Y}}_{a,3}} = {{{\mathbf{R}}}_m'}{{\mathbf{H}}_{ti}}{\mathbf{w}}{{\mathbf{w}}^H}{\mathbf{H}}_{ti}^H{{{\mathbf{R}}}_m'^H}$, ${{\mathbf{Z}}_{a,3}} = {{{\mathbf{R}}}_m'}{{\mathbf{H}}_{ti}}{\mathbf{w}}{{\mathbf{w}}^H}{{\mathbf{H}}_{tr}^H}{\mathbf{H}}_{ir}^H$, $q_{a,3} = \mathbf{w}^H\mathbf{H}_{tr}^H\mathbf{H}_{tr}\mathbf{w}$ and $\Pi_{a,3} = \mathbf{X}_{a,3}\odot\mathbf{Y}_{a,3}$. 

Then, we reshape above expressions with the SDR method by introducing a new binary vector ${\mathbf{b}} = 2{\mathbf{a}_m} - {{\mathbf{1}}_L} \in {\mathbb{R}^{L \times 1}}$, where ${\mathbf{b}} = {\left[ {{b_1},{b_2}, \ldots ,{b_L}} \right]^T}$ and ${b_l} \in \left\{ { - 1,1} \right\}$. The expression in \eqref{SI_withA} can be reformulated as 
\begin{small}\begin{align} 
&\mathbf{a}_{m}^H \Pi_{a,1}\mathbf{a}_{m} + 2\text{Re}\{\mathbf{a}_{m}^T\text{diag}\{\mathbf{Z}_{a,1}\}\}+ q_{a,1}  \nonumber \\
&={\left( {\frac{{{{\mathbf{1}}_L} + {\mathbf{b}}}}{2}} \right)^T}\Pi_{a,1}\left( {\frac{{{{\mathbf{1}}_L} + {\mathbf{b}}}}{2}} \right) + {q_{a,1}} \nonumber \\ 
&\ \ \ + 2\operatorname{Re} \left\{ {{{\left( {\frac{{{{\mathbf{1}}_L} + {\mathbf{b}}}}{2}} \right)}^T}\left( { \text{diag}\{\mathbf{Z}_{a,1}\}\}} \right)} \right\} \nonumber \\
&=\frac{1}{4}{\text{Tr}}\left( {\Pi_{a,1}{{\mathbf{b}}^T{\mathbf{b}}}} \right) + \frac{1}{4}{\text{Tr}}\left( \Pi_{a,1} \right) + \frac{1}{2}{\text{Tr}}\left( {\Pi_{a,1}{{\mathbf{1}}_L}{{\mathbf{b}}^T}} \right) + {q_{a,1}} \nonumber \\ 
&\ \ \ + \operatorname{Re} \left\{ {{{\mathbf{b}}^T}\left(\text{diag}\{\mathbf{Z}_{a,1}\}\} \right)} \right\} + \operatorname{Re} \left\{ {{\mathbf{1}}_L^T\left(\text{diag}\{\mathbf{Z}_{a,1}\}\}\right)} \right\} \nonumber \\ 
&= \frac{1}{4}\left( {{\text{Tr}}\left( {{{\mathbf{\Pi }}_{a,1}}{{\mathbf{b}}^T{\mathbf{b}}}} \right) + 2\operatorname{Re} \left\{ {{\mathbf{g}_{a,1}^T}{\mathbf{b}}} \right\}} \right) + {q_{a,1}'},
\label{SI_withg}
\end{align}\end{small}

\noindent where 
\begin{small}\begin{align*}
{\mathbf{g}_{a,1}} &= 2\text{diag}\{\mathbf{Z}_{a,1}\} + {{\mathbf{h}}_{a,1}}, \nonumber \\
{{\mathbf{h}}_{a,1}} &= {\left[ {\sum\limits_{l = 1}^L {{{\left[ {{{\mathbf{\Pi }}_{a,1}}} \right]}_{1,l}}} ;\sum\limits_{l = 1}^L {{{\left[ {{{\mathbf{\Pi }}_{a,1}}} \right]}_{2,l}}} ; \ldots ;\sum\limits_{l = 1}^L {{{\left[ {{{\mathbf{\Pi }}_{a,1}}} \right]}_{L,l}}} } \right]^T}, \nonumber \\
{q_{a,1}'} &= \frac{1}{4}{\text{Tr}}\left( {{{\mathbf{\Pi }}_{a,1}}} \right) + \operatorname{Re} \left\{ {{\mathbf{1}}_L^T\left( \text{diag}\{\mathbf{Z}_{a,1}\} \right)} \right\} + {q_{a,1}}.\nonumber
\end{align*}\end{small}

In addition, by introducing another new binary vector ${\mathbf{x}} = {\left[ {{\mathbf{b}};1} \right]^T}$, we can reshape the expression in \eqref{SI_withg} to a function of ${\mathbf{X}} = {\mathbf{x}}{{\mathbf{x}}^T}$ as 
\begin{equation}\small
\frac{1}{4}{\text{Tr}}\left( {{{{\mathbf{\Pi '}}}_{a,1}}{\mathbf{X}}} \right) + {q_{a,1}'},
\label{SI_withX}
\end{equation}
\noindent where ${{{\mathbf{\Pi '}}}_{a,1}} = \left[ \begin{gathered}
  {{\mathbf{\Pi }}_{a,1}}\quad {\mathbf{g}_{a,1}} \hfill \\
  {\mathbf{g}_{a,1}^T}\quad\ \  \ 0 \hfill \\ 
\end{gathered}  \right]$.

Then, with same process in \eqref{SI_withg} and \eqref{SI_withX}, the expressions in \eqref{Jam_Power}  and \eqref{rPower_limit} can also be reformed as follows respectively.
\begin{small}\begin{align}
\frac{1}{4}{\text{Tr}}\left( {{{{\mathbf{\Pi '}}}_{a,2}}{\mathbf{X}}} \right) + {q_{a,2}'},\\
\frac{1}{4}{\text{Tr}}\left( {{{{\mathbf{\Pi '}}}_{a,3}}{\mathbf{X}}} \right) + {q_{a,3}'},
\end{align}\end{small}

\noindent where 
\begin{small}\begin{align*}
{{{\mathbf{\Pi '}}}_{a,2}} = \left[ \begin{gathered}
  {{\mathbf{\Pi }}_{a,2}}\quad {\mathbf{g}_{a,2}} \hfill \\
  {\mathbf{g}_{a,2}^T}\quad\ \  \ 0 \hfill \\ 
\end{gathered}  \right], \ \
{{{\mathbf{\Pi '}}}_{a,3}} = \left[ \begin{gathered}
  {{\mathbf{\Pi }}_{a,3}}\quad {\mathbf{g}_{a,3}} \hfill \\
  {\mathbf{g}_{a,3}^T}\quad\ \  \ 0 \hfill \\ 
\end{gathered}  \right],
\end{align*}\end{small}
\begin{small}\begin{align*}
{\mathbf{g}_{a,2}} = -2\text{diag}\{\mathbf{Z}_{a,2}\} + {{\mathbf{h}}_{a,2}}, \  {\mathbf{g}_{a,3}} = 2\text{diag}\{\mathbf{Z}_{a,3}\} + {{\mathbf{h}}_{a,3}},\\
{{\mathbf{h}}_{a,2}} = {\left[ {\sum\limits_{l = 1}^L {{{\left[ {{{\mathbf{\Pi }}_{a,2}}} \right]}_{1,l}}} ;\sum\limits_{l = 1}^L {{{\left[ {{{\mathbf{\Pi }}_{a,2}}} \right]}_{2,l}}} ; \ldots ;\sum\limits_{l = 1}^L {{{\left[ {{{\mathbf{\Pi }}_{a,2}}} \right]}_{L,l}}} } \right]^T}, \\
{{\mathbf{h}}_{a,3}} = {\left[ {\sum\limits_{l = 1}^L {{{\left[ {{{\mathbf{\Pi }}_{a,3}}} \right]}_{1,l}}} ;\sum\limits_{l = 1}^L {{{\left[ {{{\mathbf{\Pi }}_{a,3}}} \right]}_{2,l}}} ; \ldots ;\sum\limits_{l = 1}^L {{{\left[ {{{\mathbf{\Pi }}_{a,3}}} \right]}_{L,l}}} } \right]^T}, \\
{q_{a,2}'} = \frac{1}{4}{\text{Tr}}\left( {{{\mathbf{\Pi }}_{a,2}}} \right) - \operatorname{Re} \left\{ {{\mathbf{1}}_L^T\left( \text{diag}\{\mathbf{Z}_{a,2}\} \right)} \right\} + {q_{a,2}}, \\
{q_{a,3}'} = \frac{1}{4}{\text{Tr}}\left( {{{\mathbf{\Pi }}_{a,3}}} \right) + \operatorname{Re} \left\{ {{\mathbf{1}}_L^T\left( \text{diag}\{\mathbf{Z}_{a,3}\} \right)} \right\} + {q_{a,3}}. \\
\end{align*}\end{small}

With above transformations, \eqref{P6_OF3} can then be rewritten as follows:
\begin{small}\begin{align} 
  \mathop {\max }\limits_{\mathbf{X}}&\ \ \log _2 (1+ \frac{A}{P_B\left(\frac{1}{4}{\text{Tr}}\left( {{{{\mathbf{\Pi '}}}_{a,1}}{\mathbf{X}}} \right) + {q_{a,1}'}\right) + {\sigma _r^2}}) \nonumber \\ 
 &\ - \log _2 (1+ \frac{B}{P_B\left(\frac{1}{4}{\text{Tr}}\left( {{{{\mathbf{\Pi '}}}_{a,2}}{\mathbf{X}}} \right) + {q_{a,2}'} \right) + {\sigma _e^2}})  \label{P161_OF}   \\
{\text{s.t.}} & \ \ P_B\left(\frac{1}{4}{\text{Tr}}\left( {{{{\mathbf{\Pi '}}}_{a,3}}{\mathbf{X}}} \right) + {q_{a,3}'}\right) \leq\!  {P_{th}}, \tag{\ref{P161_OF}{a}}  \label{P161_sinr}\\
& \ \ {\text{diag}}\left\{ {\mathbf{X}} \right\} = {{\mathbf{1}}_{L + 1}}, \tag{\ref{P161_OF}{b}} \label{P161_1L}\\
& \ \  \text{Rank}\left( {\mathbf{X}} \right) = 1.  \tag{\ref{P161_OF}{c}} \label{P161_rank}
\end{align}\end{small}

By introducing slack variables with the same processes in \eqref{P3_OF}-\eqref{P3_OF_v}, the objective function can then be converted to concave as  
\begin{small}\begin{align} 
 \mathop {\max }\limits_{\mathbf{X}, \tau_a, \xi_a, \varrho_a}&\ \ \log _2 \left(P_{a}\right) + \frac{Q_{a}}{P_{a}\ln{2}} \left(\tau_{a} - \overbar{\tau}_{a} \right) - \xi_{a}  \label{P_MS_mode} \\
 {\text{s.t.}} & \ \ \tau_a \geq P_B \left(\frac{1}{4}{\text{Tr}}\left( {{{{\mathbf{\Pi '}}}_{a,1}}{\mathbf{X}}} \right) + {q_{a,1}'}\right) + {\sigma _r^2},  \tag{\ref{P_MS_mode}{a}} \\
 & \ \ \xi_a \geq \log _2 (1+ \frac{B}{\varrho_a}), \tag{\ref{P_MS_mode}{b}} \\
 & \ \ \varrho_a \leq P_B \left(\frac{1}{4}{\text{Tr}}\left( {{{{\mathbf{\Pi '}}}_{a,2}}{\mathbf{X}}} \right) + {q_{a,2}'}\right) + {\sigma _e^2}, \tag{\ref{P_MS_mode}{c}} \\
& \ \  \eqref{P161_sinr} - \eqref{P161_rank},  \tag{\ref{P_MS_mode}{d}}
\end{align}\end{small}

\noindent where $P_{a} = 1 + A/\overbar{\tau}_{a}$, $Q_{a} = - A/\overbar{\tau}_{a}^2$ and $\overbar{\tau}_{a}$ is the value obtained in the last iteration.

It is worth noting that the only obstacle of convexifing \eqref{P_MS_mode} is the rank constraint in \eqref{P161_rank}, which however can be replaced with a Gaussian randomization procedure as shown in \cite{Sisai_preprint}. Therefore, we firstly solve \eqref{P_MS_mode} without \eqref{P161_rank} as a convex problem, and then extract $\mathbf{A}_m$ from $\mathbf{X}$ under rank constraint with the Gaussian randomization procedure, whose details will be illustrated in next subsection.

\subsection{Overall Algorithm and complexity analysis}
The overall optimization algorithm for secrecy capacity maximization and the Gaussian randomization procedure for the mode selection in the MS mode are summarized in Algorithm 1 and Algorithm 2, respectively.
It is worth mentioning that besides the beamforming vectors, amplitudes and phase shifts matrices, relevant slack variables will also be updated in each iteration in Algorithm 1.     

The computational complexities of solving \eqref{P_w} and \eqref{P_r} to calculate the beamforming vectors ${\mathbf{w}}$ and ${\mathbf{r}}$ are in the order of $\mathcal{O}\left(M^3\right)$ and $\mathcal{O}\left(N^3\right)$, respectively. The complexities for calculating phase shifts matrices for the ES and MS modes, with \eqref{P_ES} and \eqref{P_MS}, are both in the order of $\mathcal{O}\left(L^3\right)$ \cite{convexopt}. Furthermore, the SDR scheme and Gaussian randomization procedure for the mode selection optimization with the MS-RIS introduce the complexities of $\mathcal{O}\left( {L^{3.5}} \right)$ \cite{sdrcomplexity} and $\mathcal{O}\left( {G} \right)$, respectively. Therefore, the overall complexity order of the proposed algorithm amounts to $\mathcal{O}\left( {{N_{k}} \times \max \left\{ {M^3,N^3,{L^3}} \right\}} \right)$ for the ES mode and $\mathcal{O}\left( {{N_{k}} \times \max \left\{ {M^3,N^3,{L^{3.5}},G} \right\}} \right)$ for the MS mode, where $N_{k}$ is the number of iteration before Algorithm 1 converges to the optimal value.

\begin{algorithm}[t]
\caption{Maximizations of secrecy capacity for both the ES mode and MS mode}
\label{secrecy capacity}
\algsetup{linenosize=\footnotesize}
\begin{algorithmic}
\STATE {Initialize the beamforming vector ${\mathbf{w}}^{0}$ and ${\mathbf{r}}^{0}$, the reflecting, refracting amplitudes and phase shifts matrices ${\mathbf{R}_s ^{0}}$, ${\mathbf{T}_s ^{0}}$ of the ES mode,  the phase shifts matrices of the MS mode  ${\mathbf{R}_m ^{0}}$, ${\mathbf{T}_m ^{0}}$, as well as the mode selection matrix of the MS mode $\mathbf{A}_m^0$,
compute $C_{s}\left( {{\mathbf{w}}^0,\!{\mathbf{r}}^0,\!{{\mathbf{R}}_s^0},\!{{\mathbf{T}}_s^0}} \right)$ and $C_{s}\left( {{\mathbf{w}}^0,\!{\mathbf{r}}^0,\!{{\mathbf{R}}_m^0},\!{{\mathbf{T}}_m^0},{\mathbf{A}_m^0}} \right)$, respectively. Set iteration index $k=0$ and the accuracy for iteration $\delta$.}
\REPEAT
\STATE  
1. Given ${\mathbf{r}}^{k}, {\mathbf{R}_s^{k}}$, ${\mathbf{T}_s ^{k}}$ for the ES mode and given ${\mathbf{r}}^{k},\mathbf{A}_m^k$, ${\mathbf{R}_m^{k}}$ and ${\mathbf{T}_m ^{k}}$ for the MS mode, optimize ${\mathbf{w}}^{k+1}$  by solving problem \eqref{P_w}.\\
2. Given ${\mathbf{w}}^{k+1}$ and ${\mathbf{r}}^{k}$, optimize ${\mathbf{R} _s^{k+1}}$ and ${\mathbf{T} _s^{ k+1 }}$ by solving problem \eqref{P_ES} for the ES mode. Given $\mathbf{w}^{k+1}, {\mathbf{r}}^{k}$ and $\mathbf{A}_m^k$, optimize ${\mathbf{T} _m^{ k+1 }}$ and ${\mathbf{R} _m^{ k+1 }}$ for the MS mode, by solving problem \eqref{P_MS}. \\
3. For the MS mode, optimize $\mathbf{X}$ by solving problem \eqref{P_MS_mode} with $\mathbf{w}^{k+1},\mathbf{R}_m^{k+1}$, $\mathbf{T}_m^{k+1}$ and ${\mathbf{r}}^{k}$, and retrieve $\mathbf{A}_m^{k+1}$ from $\mathbf{X}$ using Gaussian randomization procedure in Algorithm 2. \\
4. Given ${\mathbf{w}}^{k+1}, {\mathbf{R}_s^{k+1}}$, ${\mathbf{T}_s ^{k+1}}$ for the ES mode and given ${\mathbf{w}}^{k+1},\mathbf{A}_m^{k+1}$, ${\mathbf{R}_m^{k+1}}$ and ${\mathbf{T}_m ^{k+1}}$ for the MS mode, optimize ${\mathbf{r}}^{k+1}$  by solving problem \eqref{P_r}.\\
5. Compute the secrecy capacity $C_{s}\left( {{\mathbf{w}}^{k+1}, {\mathbf{r}}^{k+1},{{\mathbf{R}}_s^{k+1}},{{\mathbf{T}}_s^{k+1}}} \right)$ for the ES mode and $C_{m}\left( {{\mathbf{w}}^{k+1}, {\mathbf{r}}^{k+1}, {{\mathbf{R}}_m^{k+1}},{{\mathbf{T}}_m^{k+1}},\mathbf{A}_m^{k+1}} \right)$ for the MS mode.\\
6. Set $k = k+1$.

\UNTIL {$\frac{{\left| {C_{s}\left( {{\mathbf{w}}^{k+1}, {\mathbf{r}}^{k+1},{{\mathbf{R}}_s^{k+1}},{{\mathbf{T}}_s^{k+1}}} \right) - C_{s}\left( {{\mathbf{w}}^{k}, {\mathbf{r}}^{k},{{\mathbf{R}}_s^{k}},{{\mathbf{T}}_s^{k}}} \right)} \right|}}{C_{s}\left( {{\mathbf{w}}^{k}, {\mathbf{r}}^{k},{{\mathbf{R}}_s^{k}},{{\mathbf{T}}_s^{k}}} \right)} \leq \delta$  or  
$\frac{{\left| {C_{m}\left( {{\mathbf{w}}^{k+1},{\mathbf{r}}^{k+1},{{\mathbf{R}}_m^{k+1}},{{\mathbf{T}}_m^{k+1}},\mathbf{A}_m^{k+1}} \right) - C_{m}\left( {{\mathbf{w}}^{k},{\mathbf{r}}^{k},{{\mathbf{R}}_m^{k}},{{\mathbf{T}}_m^{k}},\mathbf{A}_m^{k}} \right)} \right|}}{C_{m}\left( {{\mathbf{w}}^{k}, {\mathbf{r}}^{k},{{\mathbf{R}}_m^{k}},{{\mathbf{T}}_m^{k}},\mathbf{A}_m^{k}} \right)} \!\leq\! \delta$.} \hfill
\end{algorithmic}
\end{algorithm}

\begin{algorithm}[t]
\caption{Gaussian randomization procedure}
\label{randomization}
\algsetup{linenosize=\footnotesize}
\begin{algorithmic}
\STATE 1. Set a number of randomization $G$,  given the SDR solution $\mathbf{X}$. \\
\STATE 2. Generate  $\boldsymbol{\xi}_g \sim \mathcal{N}\left(\mathbf{0}, \mathbf{G} \right)$, $g=1,2,...,G$ and construct a feasible point $\tilde{\boldsymbol{x}}_{g}=\operatorname{sgn}\left( {\boldsymbol{\xi}}_g \right)$.\\
\STATE 3. Determine the optimal $g^{\star}$ by calculating 
\begin{small}\begin{align} 
 g^{\star}\! =\!  \arg \max\limits_{g=1, \ldots, G} \!  
 & \log _2 (1+ \frac{A}{\frac{1}{4}{\text{Tr}}\left( {{{{\mathbf{\Pi '}}}_{a,2}}\tilde{\boldsymbol{x}}_{g}{\tilde{\boldsymbol{x}}_{g}^T}} \right) + {q_{a,2}'} + {\sigma _r^2}}) \nonumber \\
 &- \log _2 (1+ \frac{B}{\frac{1}{4}{\text{Tr}}\left( {{{{\mathbf{\Pi '}}}_{a,1}}\tilde{\boldsymbol{x}}_{g}{\tilde{\boldsymbol{x}}_{g}^T}} \right) + {q_{a,1}'} + {\sigma _e^2}}) \nonumber
 \end{align}\end{small} 
\STATE 4. Set $\mathbf{b}=[ \boldsymbol{x}_{g^{\star}}]_{1:L}$, calculate $\mathbf{a}_m=\frac{\mathbf{b}+\mathbf{1}_L}{2}$ and \\ \ \ \  output $\mathbf{A}_m=\text{diag}\left\{ {\mathbf{a}_m} \right\}$.
\end{algorithmic}
\end{algorithm}

\section{Simulation Results}
In this section, numerical simulation results will be presented to validate the performance of the proposed optimization algorithm. The 3-dimensional positions of Alice, Eve, the first transmit antenna of Bob, the first receive antenna of Bob, and the first element at the STAR-RIS are set as $\left[ {10, - 17, 1.5} \right]$ m, $\left[ {20, 0, 1.5} \right]$ m, $\left[ {0, 0, 5} \right]$ m, $\left[ {0,0.1,5} \right]$ m and $\left[ {0.2,0,5} \right]$ m, respectively. In addition, we adopt the frequency band of 6 GHz with  $\lambda=0.05$ m, and the distance between any two adjacent elements/antennas is 0.025 m. The Rayleigh path loss exponent, Rician path loss exponent and Rician factor are $\alpha=4$, $\kappa=2.5$ and $K=3 $, respectively. The transmit power of Alice and Bob are $P_A = P_B = 10$ dBm, the noise powers at Eve and Bob are $n_e=n_r=-80$ dBm, and the power threshold $P_{th}$ in the receive antennas of Bob is set to $-60$ dBm which falls into dynamic range of most ADCs \cite{SIC_new1,ADC2}. Moreover, we set the the convergence threshold $\delta = {10^{ - 5}}$ and the length of the Gaussian randomization procedure as $G=10^{3}$.

To validate the performance of our proposed scheme, we adopt two benchmarks for comparison as follows:

\begin{enumerate}
\item \textbf{Without jamming (WOJ)}: In this scheme, Bob only receives signals from Alice with optimized receive beamfroming. We choose a set of $h_{f,n}$ and $h_{f,e}$ to set a positive secrecy capacity, thus Alice can also ensure a secure communication in this scenario.   

\item \textbf{With FD jamming (WIJ) }: In this case, we consider a traditional SIC mechanism to eliminate the SI, with which the residue SI (RSI) in the receive antennas of Bob is proportionate to the transmit power of jamming as $P_{si} = \rho P_B$. Besides, we set $\rho = -110$ dB which represents a high SIC performance \cite{RSI2}. Bob then optimizes its transmit beamforming to enhance the jamming power in Eve and optimizes its receive beamforming to enhance the received SNR from Alice.

\end{enumerate}

\begin{figure}[!t]
        \centering
        \includegraphics*[width=80mm]{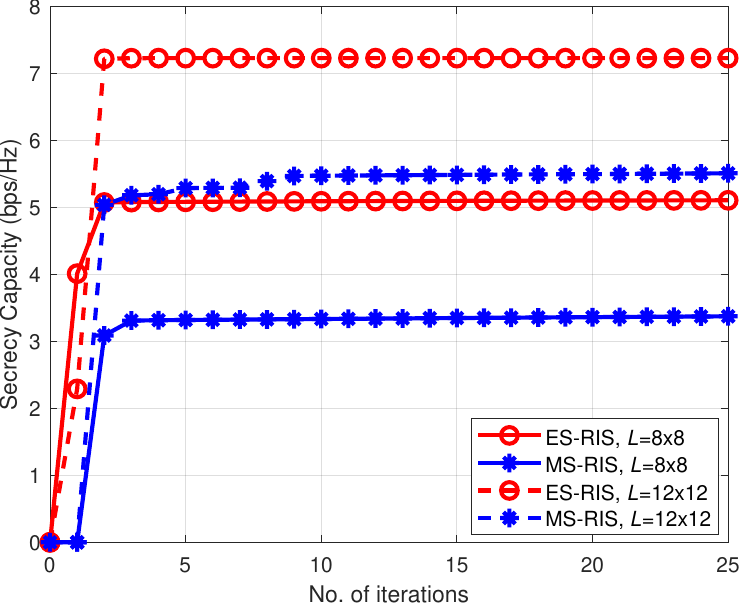}
       \caption{Convergence performance of the ES-RIS and MS-RIS, where $M=4$ and $N=2$}
        \label{fig_convegence}
\end{figure}

\begin{figure}[!t]
        \centering
        \includegraphics*[width=80mm]{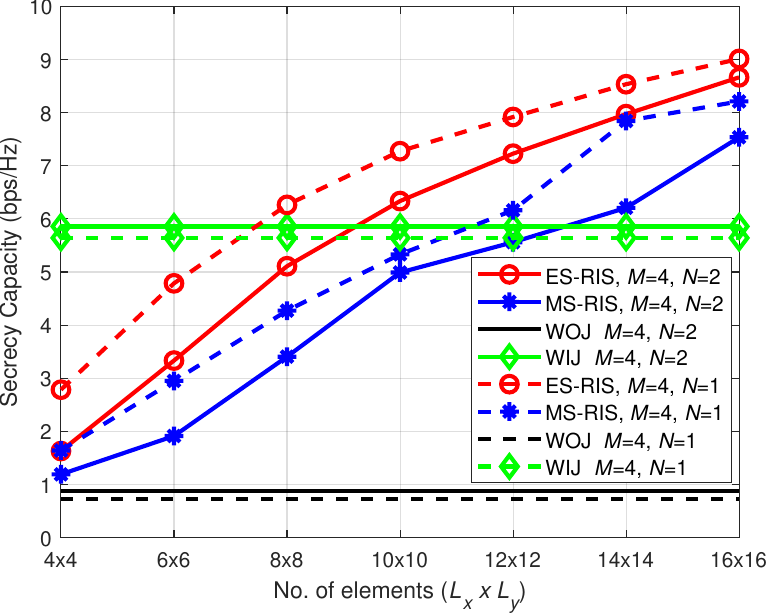}
       \caption{Secrecy Capacity Comparison with different $L$}
        \label{fig1}
\end{figure}

\begin{figure}[!t]
        \centering
        \includegraphics*[width=80mm]{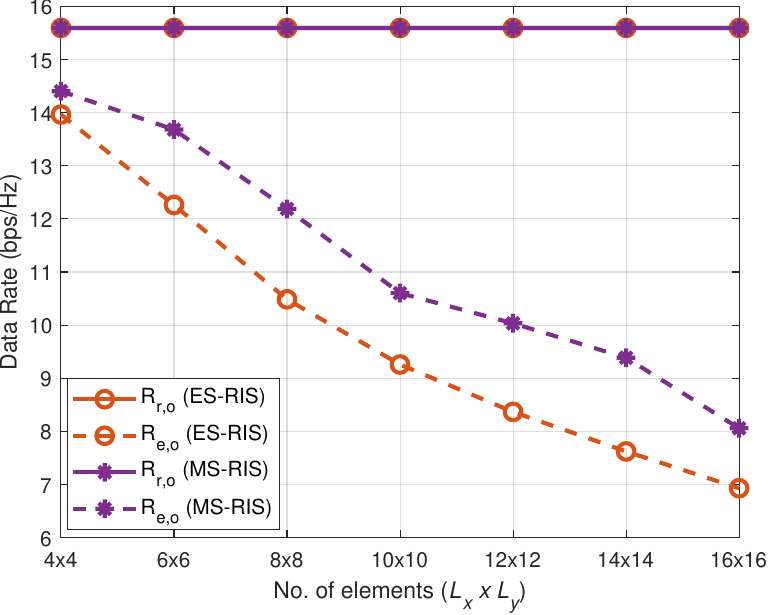}
       \caption{Data Rate of Bob and Eve with different $L$}
        \label{fig2}
\end{figure}

\begin{figure}[!t]
        \centering
        \includegraphics*[width=80mm]{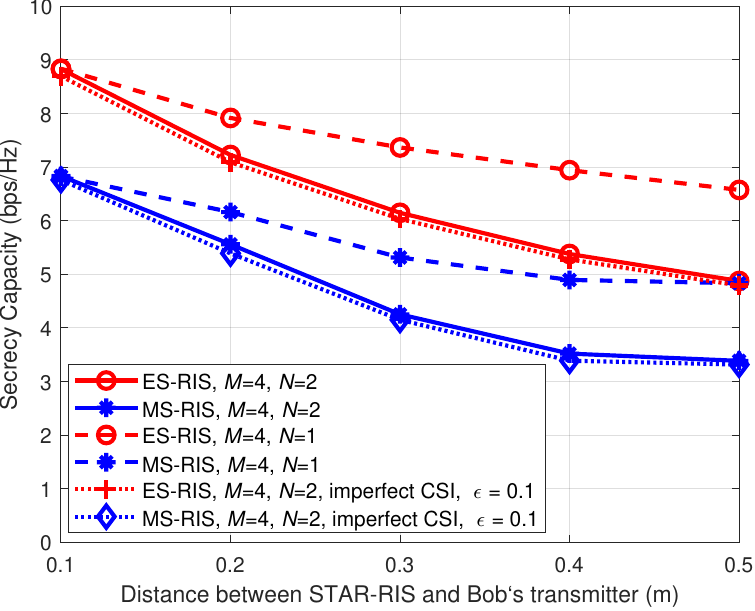}
       \caption{Secrecy capacity with different distances between the transmitter of Bob and RIS}
        \label{fig_distance}
\end{figure}

\begin{figure}[!t]
        \centering
        \includegraphics*[width=80mm]{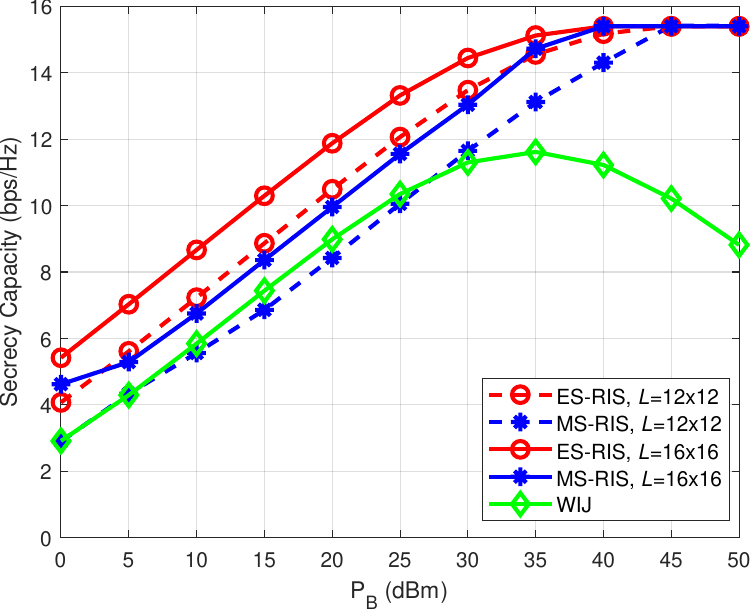}
       \caption{Secrecy capacity with different Jamming Power $P_B$}
        \label{fig_PB}
\end{figure}

The results of secrecy capacity in each iteration is shown in Fig. \ref{fig_convegence} to evaluate the convergence performance of the proposed algorithm with different number of RIS elements $L$. Through the comparison between the ES and MS modes, we can see that the MS mode requires more iterations before converging to the optimal value, especially when the number of elements is large. This can be tracked back to the fact that the proposed scheme with the MS-RIS requires an additional mode selection step when compared to that with the ES-RIS. However, it is visible from Fig. \ref{fig_convegence} that our scheme, either with the ES-RIS or MS-RIS, achieves a fast convergence with a few iterations.

The comparison of secrecy capacity with benchmarks under different numbers of $L$ is shown in Fig. \ref{fig1}. It is obvious that both the WIJ scheme and the proposed scheme, either with the ES-RIS or MS-RIS, can significantly improve the secrecy capacity compared to WOJ scheme. In addition, the secrecy capacity of our scheme with both the ES and MS modes increases with the number of elements. When the number of elements $L$ is larger than 144, the proposed algorithms with both the ES and MS modes outperform the WIJ scheme. It is because when the number of elements is large, the proposed optimization scheme can concentrate the jamming signals on Eve through the beamformed refraction of STAR-RIS, while at the same time efficiently eliminate SI as the traditional SIC. This means our STAR-RIS assisted scheme can achieve a better performance than than traditional SIC scheme with a lower cost. 

Also, we can see from Fig. \ref{fig1} that both the ES and MS modes achieve a larger secrecy capacity with number of the receive antenna of Bob $N=1$ compared to those with $N=2$. For example, when the number of elements $L =$ 144, the proposed scheme with $N=1$ achieves a secrecy capacity of 7.92 bps/Hz which is approximately 10\% higher compared to that with $N=2$. Although Bob can achieve a higher data rate with two receive antennas, as can be observed from the results of WOJ and WIJ, STAR-RIS has to reflect more power to eliminate the SI power in two antennas, which thus reduces the refraction capacity for jamming and results in a higher rate in the Eve. Furthermore, the ES mode achieves better performances of secrecy capacity than the MS mode under the same condition, which is reasonable since the domain of the optimization function for the MS mode is only a subset of that for the ES mode.  

The data rate of Bob $R_{r,o}$ and Eve $R_{e,o}$ with different number of elements are shown in Fig. \ref{fig2} with $M=4$ and $N=2$. We can see that the proposed scheme with STAR-RIS is able to cancel the SI to a sufficient low level even with only 16 elements, thus achieves a high $R_{r,o}$ for both modes. On the other hand, a larger number of elements mainly contributes to the degradation of $R_{e,o}$ by conducting a higher jamming power. We also can find that a larger $R_e$ with the MS mode under the same condition. Despite its simplicity in implementation, an element chosen for reflecting in the MS-RIS can not be further used for refraction, which results in under-utilization of RIS and a lower jamming power in the Eve.

The impact of the distance between the first antenna of STAR-RIS and that of the transmitter of Bob is shown in Fig. \ref{fig_distance}, where the number of elements $L=144$. A considerable degradation in secrecy capacity with the increase of the distance can be observed in both the ES and MS modes. For example, the secrecy capacity of the ES mode with $N=2$ drops from 8.83 to 4.87 bps/Hz when the distance increase from 0.1 to 0.5 meter. This is because a larger distance increases the path-loss of the reflecting path through RIS, thus requires a larger reflecting power ratio in the ES-RIS or a larger number of reflecting elements in the MS-RIS to tackle the SI. In addition, a larger distance between RIS and transmitter of Bob also results in a higher path loss between Bob and Eve due to the significant attenuation of $\mathbf{H}_{ti}$, which further decrease the achievable jamming power in Eve. 

We also compared the results with imperfect CSI in Fig. \ref{fig_distance}. We assume that the estimated channel between RIS and Eve is ${{{\mathbf{\bar h}}}_{ie}} = \sqrt {1 - \epsilon }  {{\mathbf{h}}_{ie}} + \sqrt {\epsilon }{{\mathbf{\tilde{h}}}_{ie}}$, where $\epsilon \! \in\! \left[ {0,1} \right]$ represents the estimation error variance and ${\mathbf{\tilde{h}}}_{ie}$ is the channel estimation error vector with ${\left[\mathbf{\tilde{h}}_{ie}\right]_{l}} \sim \mathcal{CN}\left( {0,|\left[{\mathbf{h}}_{ie}\right]_{l} |} \right)$. We can see that imperfect CSI causes a degradation on secrecy capacity compared to that with perfect CSI. 

The comparison with traditional SIC technology under different jamming power $P_B$ is shown in Fig. \ref{fig_PB} with $M=4$ and $N=2$. We can see that the secrecy capacity of traditional SIC shows an increasing trend at the beginning but eventually drops when the jamming power becomes larger. This is because that although a larger $P_B$ can decrease the rate of Eve, the RSI which depends on the SIC capacity of WIJ scheme becomes non-negligible and reduces the rate of Bob as well. On the other hand, the proposed STAR-RIS assisted schemes provide a SI cancellation which can perfectly eliminate the RSI. Therefore, the proposed algorithm with both the ES and MS modes outperforms traditional SIC with a large $P_B$ and the secrecy capacity will saturate to a upper limit when the $P_B$ is sufficiently large to decrease Eve's data rate to a value near to 0.

\section{Conclusion}
In this paper, we proposed a novel STAR-RIS assisted FD jamming scheme for secure communications, which  cancels the SI through the reflection of the RIS while simultaneously concentrate the jamming signal on the eavesdropper with the retraction function of the RIS. The joint optimization of beamforming vectors at the transmitter and receiver of FD terminal, phase shifts and amplitude for the ES-RIS, or phase shifts and mode selection for the MS-RIS were conducted to maximize the secrecy capacity. To tackle the coupling effect of multiple variables and non-convexity of the formulated problem, an alternating optimization algorithm assisted with SCA scheme was proposed. In addition, a SDR scheme with the Gaussian randomization procedure was adopted to deal with binary problem of the mode selection for the case with MS-RIS. Simulation results demonstrated the superiority of the proposed algorithm over the conventional SIC in terms of secrecy capacity. Our STAR-RIS assisted FD jamming scheme constitutes a promising candidate solution to safeguard future full duplex transmissions.

\footnotesize
\bibliographystyle{IEEEtran}
\bibliography{ref}

\begin{thebibliography}{10}
\providecommand{\url}[1]{#1}
\csname url@samestyle\endcsname
\providecommand{\newblock}{\relax}
\providecommand{\bibinfo}[2]{#2}
\providecommand{\BIBentrySTDinterwordspacing}{\spaceskip=0pt\relax}
\providecommand{\BIBentryALTinterwordstretchfactor}{4}
\providecommand{\BIBentryALTinterwordspacing}{\spaceskip=\fontdimen2\font plus
\BIBentryALTinterwordstretchfactor\fontdimen3\font minus
  \fontdimen4\font\relax}
\providecommand{\BIBforeignlanguage}[2]{{%
\expandafter\ifx\csname l@#1\endcsname\relax
\typeout{** WARNING: IEEEtran.bst: No hyphenation pattern has been}%
\typeout{** loaded for the language `#1'. Using the pattern for}%
\typeout{** the default language instead.}%
\else
\language=\csname l@#1\endcsname
\fi
#2}}
\providecommand{\BIBdecl}{\relax}
\BIBdecl

\bibitem{Security1}
V.-L. Nguyen, P.-C. Lin, B.-C. Cheng, R.-H. Hwang, and Y.-D. Lin, ``Security
  and privacy for {6G}: A survey on prospective technologies and challenges,''
  \emph{IEEE Communications Surveys \& Tutorials}, vol.~23, no.~4, pp.
  2384--2428, 4th Quart. 2021.

\bibitem{Security2}
N.~Neshenko, E.~Bou-Harb, J.~Crichigno, G.~Kaddoum, and N.~Ghani,
  ``Demystifying {IoT} security: An exhaustive survey on {IoT} vulnerabilities
  and a first empirical look on internet-scale {IoT} exploitations,''
  \emph{IEEE Communications Surveys \& Tutorials}, vol.~21, no.~3, pp.
  2702--2733, 3rd Quart. 2019.

\bibitem{PLS1}
A.~Chorti, A.~N. Barreto, S.~Köpsell, M.~Zoli, M.~Chafii, P.~Sehier,
  G.~Fettweis, and H.~V. Poor, ``Context-aware security for {6G} wireless: The
  role of physical layer security,'' \emph{IEEE Communications Standards
  Magazine}, vol.~6, no.~1, pp. 102--108, Mar. 2022.

\bibitem{PLS_survey}
P.~Angueira, I.~Val, J.~Montalban, O.~Seijo, E.~Iradier, P.~S. Fontaneda,
  L.~Fanari, and A.~Arriola, ``A survey of physical layer techniques for secure
  wireless communications in industry,'' \emph{IEEE Communications Surveys \&
  Tutorials}, vol.~24, no.~2, pp. 810--838, 2nd Quart. 2022.

\bibitem{gaojiepls}
G.~Chen, Y.~Gong, P.~Xiao, and J.~A. Chambers, ``Physical layer network
  security in the full-duplex relay system,'' \emph{IEEE Transactions on
  Information Forensics and Security}, vol.~10, no.~3, pp. 574--583, Mar. 2015.

\bibitem{FD1}
S.~Zhao, J.~Liu, Y.~Shen, X.~Jiang, and N.~Shiratori, ``Secure beamforming for
  full-duplex {MIMO} two-way untrusted relay systems,'' \emph{IEEE Transactions
  on Information Forensics and Security}, vol.~15, pp. 3775--3790, Jun. 2020.

\bibitem{FD2}
Z.~Kong, S.~Yang, D.~Wang, and L.~Hanzo, ``Robust beamforming and jamming for
  enhancing the physical layer security of full duplex radios,'' \emph{IEEE
  Transactions on Information Forensics and Security}, vol.~14, no.~12, pp.
  3151--3159, Dec. 2019.

\bibitem{FD3}
R.~Sohrabi, Q.~Zhu, and Y.~Hua, ``Secrecy analyses of a full-duplex {MIMOME}
  network,'' \emph{IEEE Transactions on Signal Processing}, vol.~67, no.~23,
  pp. 5968--5982, Dec. 2019.

\bibitem{FD_SI}
F.~Jameel, S.~Wyne, G.~Kaddoum, and T.~Q. Duong, ``A comprehensive survey on
  cooperative relaying and jamming strategies for physical layer security,''
  \emph{IEEE Communications Surveys \& Tutorials}, vol.~21, no.~3, pp.
  2734--2771, 3rd Quart. 2019.

\bibitem{SIC_new1}
M.~A. Islam, G.~C. Alexandropoulos, and B.~Smida, ``Joint analog and digital
  transceiver design for wideband full duplex {MIMO} systems,'' \emph{IEEE
  Transactions on Wireless Communications}, vol.~21, no.~11, pp. 9729--9743,
  Nov. 2022.

\bibitem{BeamBased}
A.~T. Le, L.~C. Tran, X.~Huang, and Y.~J. Guo, ``Beam-based analog
  self-interference cancellation in full-duplex {MIMO} systems,'' \emph{IEEE
  Transactions on Wireless Communications}, vol.~19, no.~4, pp. 2460--2471,
  Apr. 2020.

\bibitem{SIC_new3}
M.~Ghoraishi, W.~Jiang, P.~Xiao, and R.~Tafazolli, ``Subband approach for
  wideband self-interference cancellation in full-duplex transceiver,'' in
  \emph{2015 International Wireless Communications and Mobile Computing
  Conference (IWCMC)}, Aug 2015, pp. 1139--1143.

\bibitem{SIC_new4}
D.~Liang, P.~Xiao, G.~Chen, M.~Ghoraishi, and R.~Tafazolli, ``Digital
  self-interference cancellation for full-duplex mimo systems,'' in \emph{2015
  International Wireless Communications and Mobile Computing Conference
  (IWCMC)}, Aug 2015, pp. 403--407.

\bibitem{HYBF2}
K.~Satyanarayana, M.~El-Hajjar, P.-H. Kuo, A.~Mourad, and L.~Hanzo, ``Hybrid
  beamforming design for full-duplex millimeter wave communication,''
  \emph{IEEE Transactions on Vehicular Technology}, vol.~68, no.~2, pp.
  1394--1404, Feb. 2019.

\bibitem{SIC_new2}
R.~Askar, J.~Chung, Z.~Guo, H.~Ko, W.~Keusgen, and T.~Haustein, ``Interference
  handling challenges toward full duplex evolution in {5G} and beyond cellular
  networks,'' \emph{IEEE Wireless Communications}, vol.~28, no.~1, pp. 51--59,
  Feb. 2021.

\bibitem{Cognitive}
M.~A. ElMossallamy, H.~Zhang, L.~Song, K.~G. Seddik, Z.~Han, and G.~Y. Li,
  ``Reconfigurable intelligent surfaces for wireless communications:
  Principles, challenges, and opportunities,'' \emph{IEEE Transactions on
  Cognitive Communications and Networking}, vol.~6, no.~3, pp. 990--1002, Sep.
  2020.

\bibitem{magRIS}
C.~Pan, H.~Ren, K.~Wang, J.~F. Kolb, M.~Elkashlan, M.~Chen, M.~Di~Renzo,
  Y.~Hao, J.~Wang, A.~L. Swindlehurst, X.~You, and L.~Hanzo, ``Reconfigurable
  intelligent surfaces for {6G} systems: Principles, applications, and research
  directions,'' \emph{IEEE Communications Magazine}, vol.~59, no.~6, pp.
  14--20, Jun. 2021.

\bibitem{RIS-sec1}
Z.~Chu, W.~Hao, P.~Xiao, D.~Mi, Z.~Liu, M.~Khalily, J.~R. Kelly, and A.~P.
  Feresidis, ``Secrecy rate optimization for intelligent reflecting surface
  assisted {MIMO} system,'' \emph{IEEE Transactions on Information Forensics
  and Security}, vol.~16, pp. 1655--1669, 2021.

\bibitem{RIS-sec2}
J.~Zhang, H.~Du, Q.~Sun, B.~Ai, and D.~W.~K. Ng, ``Physical layer security
  enhancement with reconfigurable intelligent surface-aided networks,''
  \emph{IEEE Transactions on Information Forensics and Security}, vol.~16, pp.
  3480--3495, May 2021.

\bibitem{RIS-FD-sec1}
Y.~Ge and J.~Fan, ``Robust secure beamforming for intelligent reflecting
  surface assisted full-duplex {MISO} systems,'' \emph{IEEE Transactions on
  Information Forensics and Security}, vol.~17, pp. 253--264, 2022.

\bibitem{RIS-FD-sec2}
Y.~Jin, R.~Guo, L.~Zhou, and Z.~Hu, ``Secure beamforming for {IRS}-assisted
  nonlinear {SWIPT} systems with full-duplex user,'' \emph{IEEE Communications
  Letters}, vol.~26, no.~7, pp. 1494--1498, Jul. 2022.

\bibitem{NTT}
N.~DOCOMO., ``{DOCOMO} conducts world’s first successful trial of transparent
  dynamic metasurface,''
  \url{https://www.nttdocomo.co.jp/english/info/media_center/pr/2020/0117_00.html}.

\bibitem{STAR-1}
Y.~Liu, X.~Mu, J.~Xu, R.~Schober, Y.~Hao, H.~V. Poor, and L.~Hanzo, ``{STAR}:
  Simultaneous transmission and reflection for 360° coverage by intelligent
  surfaces,'' \emph{IEEE Wireless Communications}, vol.~28, no.~6, pp.
  102--109, Dec. 2021.

\bibitem{IOS_cov}
S.~Zhang, H.~Zhang, B.~Di, Y.~Tan, M.~Di~Renzo, Z.~Han, H.~Vincent~Poor, and
  L.~Song, ``Intelligent omni-surfaces: Ubiquitous wireless transmission by
  reflective-refractive metasurfaces,'' \emph{IEEE Transactions on Wireless
  Communications}, vol.~21, no.~1, pp. 219--233, Jan. 2022.

\bibitem{Sisai_preprint}
S.~Fang, G.~Chen, P.~Xiao, K.-K. Wong, and R.~Tafazolli, ``Intelligent omni
  surface-assisted self-interference cancellation for full-duplex {MISO}
  system,'' \emph{arXiv preprint arXiv:2208.06457}, 2022.

\bibitem{STAR-2}
X.~Mu, Y.~Liu, L.~Guo, J.~Lin, and R.~Schober, ``Simultaneously transmitting
  and reflecting {(STAR)} {RIS} aided wireless communications,'' \emph{IEEE
  Transactions on Wireless Communications}, vol.~21, no.~5, pp. 3083--3098, May
  2022.

\bibitem{CSI1}
Q.~Wu and R.~Zhang, ``Towards smart and reconfigurable environment: Intelligent
  reflecting surface aided wireless network,'' \emph{IEEE Communications
  Magazine}, vol.~58, no.~1, pp. 106--112, Jan. 2020.

\bibitem{CSI2}
Z.-Q. He and X.~Yuan, ``Cascaded channel estimation for large intelligent
  metasurface assisted massive {MIMO},'' \emph{IEEE Wireless Communications
  Letters}, vol.~9, no.~2, pp. 210--214, Feb. 2020.

\bibitem{ADC2}
A.~Sabharwal, P.~Schniter, D.~Guo, D.~W. Bliss, S.~Rangarajan, and R.~Wichman,
  ``In-band full-duplex wireless: Challenges and opportunities,'' \emph{IEEE
  Journal on Selected Areas in Communications}, vol.~32, no.~9, pp. 1637--1652,
  Sep. 2014.

\bibitem{Convex_ineq}
Z.~Sheng, H.~D. Tuan, A.~A. Nasir, T.~Q. Duong, and H.~V. Poor, ``Power
  allocation for energy efficiency and secrecy of wireless interference
  networks,'' \emph{IEEE Transactions on Wireless Communications}, vol.~17,
  no.~6, pp. 3737--3751, Jun. 2018.

\bibitem{Matrix_Analysis}
X.~D. {Zhang}, \emph{{Matrix Analysis and Applications}}, 2017.

\bibitem{convexopt}
S.~Boyd and L.~Vandenberghe, \emph{Convex Optimization}.\hskip 1em plus 0.5em
  minus 0.4em\relax Cambridge University Press, 2004.

\bibitem{sdrcomplexity}
Z.-q. Luo, W.-k. Ma, A.~M.-c. So, Y.~Ye, and S.~Zhang, ``Semidefinite
  relaxation of quadratic optimization problems,'' \emph{IEEE Signal Processing
  Magazine}, vol.~27, no.~3, pp. 20--34, May 2010.

\bibitem{RSI2}
Z.~Xiao and Y.~Zeng, ``Waveform design and performance analysis for full-duplex
  integrated sensing and communication,'' \emph{IEEE Journal on Selected Areas
  in Comm.}, vol.~40, no.~6, pp. 1823--1837, Jun. 2022.

\end{thebibliography}
\end{document}